\documentclass[a4paper,11pt]{article}
\pdfoutput=1
\usepackage[utf8]{inputenc}
\usepackage{jcappub}

\begin{document}

\begin{flushright}
	PI/UAN-2020-667FT\\
    HIP-2020-23/TH
\end{flushright}
\title{\centering Chiral gravitational waves and primordial black holes in\\UV-protected Natural Inflation}
\author[a]{Juan P. Beltr\'an Almeida,}
\author[b]{Nicol\'as Bernal,}
\author[c]{\\Dario Bettoni,}
\author[d,e]{Javier Rubio}
\affiliation[a]{Universidad Nacional de Colombia, Facultad de Ciencias, Departamento de F\'isica,\\Av. Cra 30 \# 45-03, Bogot\'a, Colombia}
\affiliation[b]{Centro de Investigaciones, Universidad Antonio Nari\~no, \\ Cra 3 Este \# 47A-15, Bogot\'a, Colombia}
\affiliation[c]{Departamento de F\'isica Fundamental and IUFFyM,\\ 
Universidad de Salamanca, E 37008 Salamanca, Spain}
\affiliation[d]{Department of Physics and Helsinki Institute of Physics, \\  PL 64, FI-00014 University of Helsinki, Finland} 
\affiliation[e]{Centro de Astrof{\'i}sica e Gravita\c c\~ao-CENTRA,\\
Departamento de F{\'i}sica, Instituto Superior T{\'e}cnico-IST,\\
Universidade de Lisboa-UL, Avenida Rovisco Pais 1, 1049-001, Lisboa, Portugal}
\emailAdd{jubeltrana@unal.edu.co}
\emailAdd{nicolas.bernal@uan.edu.co}
\emailAdd{bettoni@usal.es}
\emailAdd{javier.rubio@tecnico.ulisboa.pt}

\abstract{
We consider an UV-protected Natural Inflation scenario involving Chern-Simons-like interactions between the inflaton and some beyond the Standard Model gauge fields. The accelerated expansion of the Universe is supported by a combination of a gravitationally-enhanced friction sensitive to the scale of inflation and quantum friction effects associated with the explosive production of gauge fluctuations.
The synergy of these two velocity-restraining mechanisms allows for: $i)$ Natural Inflation potentials involving only sub-Planckian coupling constants, $ii)$ the generation of a dark matter component in the form of primordial black holes, and $iii)$ a potentially observable background of chiral gravitational waves at small scales.
}
\maketitle

\section{Introduction and discussion}

Ground-breaking experiments such as WMAP \cite{Hinshaw:2012aka} and Planck \cite{Akrami:2018odb} have consolidated inflation as the standard mechanism for the generation of the primordial density perturbations seeding structure formation.  To match observations, a canonically normalized inflaton field must be endowed with a sufficiently flat potential, protected from quantum corrections by global symmetries such as dilatation invariance \cite{GarciaBellido:2011de,Bezrukov:2012hx,Rubio:2014wta,Burgess:2014lza,Bezrukov:2017dyv,Rubio:2020zht} or shift symmetry \cite{Broy:2015qna,Finelli:2018upr}. The latter possibility is typically realized in \textit{Natural Inflation} scenarios \cite{Freese:1990rb, Adams:1992bn,Freese:2014nla}. In this type of settings, the inflaton field is identified with a pseudo-Nambu-Goldstone boson or axion, which, as happens in the Peccei-Quinn mechanism \cite{Peccei:1977hh}, acquires a symmetry breaking potential at a scale $f$ via instanton effects. 

One of the main difficulties of \textit{Natural Inflation} is associated to its ultraviolet (UV) completion. In particular, the super-Planckian values of $f$ needed to support a sufficiently long period of inflation are in tension with the usual expectations of controlled string compactifications~\cite{Banks:2003sx, Svrcek:2006yi} and the weak gravity conjecture \cite{ArkaniHamed:2006dz}. Several ways of reconciling inflation with sub-Planckian coupling constants have been proposed in the literature (see e.g. Refs.~\cite{Kawasaki:2000yn,ArkaniHamed:2003wu,Kim:2004rp,Dimopoulos:2005ac,Grimm:2007hs,McAllister:2008hb,Kaloper:2008fb,Czerny:2014wza,Choi:2014rja,Kappl:2014lra, Carta:2020oci}). An interesting possibility not requiring the inclusion additional degrees of freedom is to introduce a non-minimal coupling between the Einstein tensor and the inflaton kinetic term. In this so-called \textit{UV-protected Natural Inflation} scenario  \cite{Germani:2010hd,Germani:2011ua}, the inflaton friction is \textit{gravitationally enhanced}, allowing for an accelerated expansion even in steep potentials with sub-Planckian coupling constants.

In this work we re-examine the above \textit{UV-protected} scenario in the presence of parity-violating Chern-Simons interactions~\cite{Anber:2009ua,Anber:2012du}.\footnote{Due to the many appealing features of Chern-Simons interactions---such as the sourcing of chiral gravitational waves, the appearance of parity breaking and anisotropic patterns in the cosmic microwave background, the possible connections with magnetogenesis and the production of primordial black holes---, these kind of scenarios have been enthusiastically studied in the literature. An incomplete list includes, for instance, Refs.~\cite{Anber:2006xt, Durrer:2010mq, Barnaby:2010vf, Barnaby:2011vw, Barnaby:2011qe, Barnaby:2012tk, Adshead:2012kp, Dimopoulos:2012av, Meerburg:2012id, Barnaby:2012xt, Linde:2012bt, Cook:2013xea, Shiraishi:2013kxa, Fleury:2014qfa, Caprini:2014mja, Ferreira:2014zia, Bartolo:2014hwa, Bartolo:2015dga, Namba:2015gja, Ferreira:2015omg, Domcke:2016bkh, Peloso:2016gqs,  Shiraishi:2016yun, Garcia-Bellido:2016dkw, Obata:2016tmo,Obata:2016xcr,Jimenez:2017cdr, Caprini:2017vnn,Adshead:2018doq,Adshead:2019igv,Adshead:2019lbr, Ozsoy:2020ccy,Watanabe:2020ctz}.} The evolution of the inflaton field during inflation translates into the spontaneous symmetry breaking of conformal symmetry and the quantum generation of gauge fluctuations that tend to dissipate the background energy density. This Schwinger-like mechanism, clearly reminiscent of warm inflation scenarios~\cite{Berera:1995ie}, slows down the evolution of the inflaton field even when the  \textit{gravitationally-enhanced friction} ceases to be efficient. 

We perform here a detailed comparison of the above model's predictions with present and future data sets, highlighting the similarities and differences with previous studies in the literature. At early times, the evolution of the system is dominated by the non-minimal derivative coupling to gravity. The inflaton velocity is correspondingly small and the gauge fluctuations are completely subdominant. This translates into an almost flat spectrum of primordial density perturbations in perfect agreement with cosmic microwave background (CMB) observations for an extensive range of sub-Planckian axion constants. The rise of the field velocity as inflation proceeds leads to the exponential growth of the vector contributions, which source subsequently the scalar and tensor perturbations. The enhancement of scalar perturbations with respect to their CMB values allows for the potential formation of primordial black holes (PBH) at sub-CMB scales. We argue that a proper choice of the model parameters can easily accommodate present observational constraints on these appealing dark matter candidates. Finally, we confront the scenario with the sensitivity of future gravitational wave (GW) interferometers.  In particular, we show that the late-time amplification of the tensor power spectrum during the axial regime allows to obtain an observable GW signal which is both non-Gaussian and maximally chiral.

\section{The model}\label{sec:model}

We consider an UV-protected Natural Inflation scenario involving a pseudo-scalar inflaton field $\phi$ interacting with ${\cal N}$ gauge fields $A^a_\mu$ via Chern-Simons interactions. The action of the model takes the form 
\begin{equation}\label{KHAS}
S = \int {\rm d}^{4}x \sqrt{-g}\left[ \frac{M_P^2}{2}R  -\frac{1}{2}\left(g^{\mu \nu} - \frac{1}{M^2}G^{\mu \nu} \right) \nabla_{\mu}\phi \nabla_{\nu}\phi- V(\phi) -\frac{1}{4} F^{\mu \nu}_a F^a_{\mu \nu}   - \frac{\alpha}{4}\frac{\phi}{f} \tilde{F}_a^{\mu \nu} F^a_{\mu \nu}  \right],
\end{equation}
with $G^{\mu\nu} \equiv R^{\mu\nu} - \frac{1}{2}R\,g^{\mu\nu}$ the Einstein tensor, the index $a$ ranging from 1 to ${\cal N}$, and $F^a_{\mu\nu}\equiv\partial_\mu A^a_\nu - \partial_\nu A^a_\mu$ and $\sqrt{-g}\,\widetilde{F}^a_{\mu\nu}=\frac{1}{2}\epsilon^{\mu\nu\sigma\tau}F^a_{\sigma\tau}$ the field strength associated with each gauge field and its dual. Here $M$ is a constant with the dimension of mass, $\epsilon^{0123}=1$ and 
\begin{equation}\label{eq:pot}
V(\phi) = \Lambda^4 \left[ 1 + \cos \left(\frac{\phi}{f}\right) \right]
\end{equation}
an axion-like cosine potential.  This type of action may appear naturally in scenarios involving several axion-like fields  {interacting non-minimally with an extended gauge sector of strength ${\cal F}_{\mu\nu}^a\neq F_{\mu\nu}^a$ beyond \eqref{KHAS}}.  In particular, if one of  {the additional gauge fields in ${\cal F}_{\mu\nu}^a$} enters into a strong coupling regime at an energy scale $\Lambda$, it may generate a periodic potential with amplitude $\Lambda^4$ and an effective Chern-Simons interaction {$\alpha\, \phi \,\tilde{F}_a^{\mu \nu} F^a_{\mu \nu}$ with} $\alpha\gg 1$ (see, for instance, Ref.~\cite{Choi:1985bz}). The resulting action is technically natural since the shift symmetry on $\phi$ is effectively restored in the $\Lambda\to 0$ limit. This important feature is respected by the non-minimal coupling $G^{\mu\nu}\nabla_\mu\phi \nabla_\mu \phi$, which, in spite of involving a higher number of derivatives, does not introduce additional degrees of freedom beyond those originally present in the theory~\cite{Germani:2010gm}.  

In an isotropic and flat Friedmann-Lema\^itre-Robertson-Walker spacetime ${\rm d}s^2 = -{\rm d}t^2 + a^2(t)\,\delta_{ij}\,{\rm d}x^i {\rm d}x^j$ with scale factor $a(t)$, the Friedmann equations following from the variation of the action~\eqref{KHAS} with respect to the metric take the form~\cite{Sushkov:2009hk,Gumjudpai:2015vio}
\begin{eqnarray}
 && H^2  =  \frac{1}{  3M_P^2}\left[\frac{1}{2}\dot{\phi}^2\left(1 + 9\frac{H^2}{M^2} \right) + V(\phi) + \frac{ {1}}{2}\left( \langle \vec{E_a^2}\rangle + \langle \vec{B_a^2}\rangle \right)\right],  \label{Friedman1}\\
&& \dot{{H}}   = - \frac{1}{2M_P^2}  \left[\left(1+ 3 \frac{H^2}{M^2}\right) \dot{\phi}^2 -\frac{1}{M^2}\frac{d(H \dot\phi^2)}{dt} \right]  - \frac{1}{ 3 M_P^2}  \left( \langle \vec{E_a^2}\rangle + \langle \vec{B_a^2}\rangle \right), \label{Friedman2}
\end{eqnarray}
with the dots denoting derivatives with respect to the coordinate time $t$. Here we have particularized to the Coulomb gauge $A^a_0=\vec \nabla\cdot\vec A^a=0 $, used the standard electromagnetic notation to denote the ``electric'' ($\vec E^a$) and ``magnetic'' ($\vec B^a$) gauge field components and assumed a mean field approximation to account for the backreaction of the associated fluctuations, with  the brackets denoting \textit{quantum expectation values} (for details cf.~Section~\ref{sec:vecBR}). The Friedmann equations \eqref{Friedman1} and \eqref{Friedman2} are supplemented by the equations of motion for the inflaton and vector fields, namely
\begin{equation}
\frac{1}{a^3}\frac{d}{dt} \left[a^3\dot\phi\, K  \right] + V_{\phi} = \frac{  \alpha}{f}  \langle  \vec{E}_a \cdot  \vec{B}_a  \rangle  \,, 
\hspace{10mm} \vec{A}^a{}'' - \nabla^2 \vec{A}^a =\frac{\alpha\,}{f}\,  \phi' \,{\vec{\nabla}} \times \vec{A}^a  \,, \label{eq:phi}
\end{equation}
with the primes denoting derivatives with respect to the conformal time $\tau\equiv \int dt/a$,  $V_{\,\phi}\equiv \partial V/\partial\phi$ and
\begin{equation}\label{Kdef}
K \equiv 1+3\frac{H^2}{M^2}\, . 
\end{equation}

\subsection{Vector field production and backreaction} \label{sec:vecBR}

To determine the averages in Eqs.~\eqref{Friedman1}-\eqref{eq:phi}, let us perform a canonical quantization of the gauge fields, namely
\begin{equation}\label{eq:Fourier}
\vec{A}^a\left(\tau,\vec{x}\right) = \sum_{\lambda=\pm}
\int \frac{d^3 \vec{k}}{\left(2\pi\right)^{3/2}}\left[
A^a_\lambda (\tau,\vec{k})\,\vec{\epsilon}_\lambda(\vec{k})\,
\hat{a}_\lambda(\vec{k})\, e^{i\,\vec{k}\cdot\vec{x}} +\textrm{H.c.}\right],
\end{equation}
with $A^a_\lambda$ the mode functions associated with the two helicities $\lambda=\pm 1$. Here  $\hat{a}_\lambda (\vec{k})$ and $\hat{a}_\lambda^\dagger (\vec{k})$ stand for the usual annihilation and creation operators satisfying the canonical commutation relations 
$\big[\hat{a}_\lambda (\vec{k}),\hat{a}_{\lambda'}^\dagger(\vec{k}')\big]
= \delta_{\lambda\lambda'}\,\delta^{(3)}(\vec{k}-\vec{k}')$ and  $\vec{\epsilon}_\pm$ is an orthonormal basis in the complex vector space perpendicular to the momentum, i.e.
\begin{equation}
	\vec{\epsilon}_\lambda(\vec{k}) \cdot \vec{\epsilon}_{\lambda'}^{\,*}(\vec{k})
= \delta_{\lambda\lambda'} \,, \hspace{10mm}
\vec{\epsilon}_\lambda(\vec{k}) \cdot \vec{k} = 0 \,, \hspace{10mm}
i {\vec{k}}\times\vec{\epsilon}_\lambda(\vec{k}) =
\lambda\, k\, \vec{\epsilon}_\lambda(\vec{k}) \,, \hspace{10mm}  k = \vert \vec{k}\vert\,.
\end{equation}
Plugging the expansion~\eqref{eq:Fourier} into the corresponding field equation in~\eqref{eq:phi} and taking into account that $\phi'= - \dot \phi/(H\tau) $, we get
\begin{equation} \label{eqApp}
A_\pm^a{}''  + \left[ k\left(k  \mp \frac{2 \xi  }{\vert \tau\vert } \right)  \right]A^a_{\pm} = 0\,,
\end{equation}
where we have used the background isotropy to relabel the modes $A^a_\pm (\tau,\vec{k})$ as functions $A^a_\pm$ depending only on the absolute value of the momenta.  The \textit{instability parameter} 
\begin{equation}\label{eq:inst_par}
 \xi \equiv \frac{\alpha\, \dot{\phi}}{2\,f\, H}
\end{equation} 
grows adiabatically with time, meaning that its value should be understood as the one acquired when the mode under consideration crosses the horizon. 

The mode equation \eqref{eqApp} displays the parity-violating nature of the system. While negative-helicity modes $A^a_{-}$ experience just a shift in their dispersion relation for positive $\xi$ (i.e. for $\dot \phi>0$), the positive-helicity modes $A^a_{+}$ become tachyonically unstable for 
\begin{equation}\label{kcrit}
 k\,\, <\,\,k_{\rm cr}\equiv\frac{2\,\xi}{\vert\tau\vert}\,.
\end{equation}
This qualitative behavior is consistent with the exact solutions of Eq.~\eqref{eqApp} satisfying the Bunch-Davies boundary condition $\lim_{-k\tau\rightarrow \infty} A^a_\pm \left(\tau\right) = e^{-ik\tau}/\sqrt{2k} $, namely  \begin{equation} \label{Apmsol}
 A^a_{\pm} (x)=\frac{e^{\pm \frac{\pi}{2} \xi }}{\sqrt{2\, k}} W_{\pm i\xi,\, \frac12}(2i\, x) \simeq \frac{1}{\sqrt{2k}} \left( \frac{k}{2\,\xi\, a\, H}\right)^{1/4} e^{\pi\, \xi - 2 \sqrt{2\xi \,k/(aH)}}\,,
\end{equation}
with $x\equiv -k\, \tau$, $W$ the regular Whittaker function, and the right-hand side approximation corresponding to $k\, \ll k_{\rm cr}$. Neglecting the vanishing contribution of negative-helicity modes and taking into account the de Sitter scale factor $a=-1/(H\tau)$, we can express the quantum averages in Eqs.~\eqref{Friedman1}, \eqref{Friedman2} and~\eqref{eq:phi} as \cite{Anber:2009ua}
\begin{align}
\langle \vec{E}_a^2\rangle  &\simeq  \frac{1}{a^4} \int \frac{d^3\vec k}{\left(2\pi\right)^3}
\left|\frac{\partial}{\partial \tau} A^a_+\right|^2=  I_1\times {\cal N} \, H^4 \,\frac{e^{2\pi\xi}}{\xi^3} \,, \label{eq:E2}
\\ 
\langle \vec{B}_a^2\rangle &\simeq  \frac{1}{a^4} \int \frac{d^3\vec k}{\left(2\pi\right)^3}
\, k^2 \left|A^a_+\right|^2\simeq  I_2\times  {\cal N} \, H^4\,\frac{e^{2\pi\xi}}{\xi^5} \,,  \label{eq:B2} \\
\langle \vec{E}_a\cdot \vec{B}_a\rangle & = -\frac{1}{2 a^4} \int \frac{d^3 \vec k}{(2\pi)^3} \,k\, \frac{\partial}{\partial \tau} |A^a_{+}|^2  = - I_3  \times  {\cal N}\, H^4\, \frac{e^{2\pi\xi}}{\xi^4} \,.\label{eq:EB}
\end{align}
Here \cite{Jimenez:2017cdr,Ballardini:2019rqh}
\begin{align}
\label{eq:integrals}
I_{1} & = \frac{\xi^3}{4\pi^2}\, e^{-\pi\,\xi}
\int_{0}^{x_{\rm c}} dx\,x^3 \left|\frac{\partial}{\partial x}
W_{i\xi,1/2}\left(2ix\right)\right|^2 \underset{\xi \gg 1}{\simeq}  2.6\times 10^{-4
} \,,
\\ 
I_{2} & = \frac{\xi^5}{4\pi^2}\, e^{-\pi\,\xi}
\int_{0}^{x_{\rm c}} dx\,x^3 \left|W_{i\xi,1/2}\left(2ix\right)\right|^2 \underset{\xi \gg 1}{\simeq} 3\times 10^{-4}   \,,
\\ 
I_{3}& = -\frac{\xi^4}{8\pi^2}\, e^{-\pi\,\xi}
\int_{0}^{x_{\rm c}} dx\,x^3\,\frac{\partial}{\partial x}
\left|W_{i\xi,1/2}\left(2ix\right)\right|^2 \underset{\xi \gg 1}{\simeq} 2.6\times 10^{-4}  \,,
\end{align}
are some appropriately defined functions that for $\xi\gtrsim 3$ become insensitive to the precise choice of the cutoff $x_c$ regularizing the infinite contribution of high-frequency modes. Note that while the growth of the modes $A^a_+$ takes its maximum value at the momenta $k \simeq k_{\rm cr}/2$ maximizing the tachyonic  frequency in Eq.~\eqref{eqApp}, the larger contribution to  $\langle \vec{E}_a\cdot \vec{B}_a\rangle$ comes from scales $k\simeq k_{\rm cr}$, due to the additional $\tau$ derivative and the approximate $k^4$ dependence in Eq.~\eqref{eq:EB}. 

\section{Primordial power spectra}\label{sec:spectra}

Having determined the influence of the gauge fields in the background equations of motion, we calculate now the primordial power spectra of scalar and tensor perturbations.  The scenario with $K=1$ ($M\to \infty$) was considered in Ref.~\cite{Anber:2009ua}, where it was shown that the generated scalar perturbations can only reproduce the observed amplitude and Gaussianity of CMB perturbations if the pseudoscalar inflaton field $\phi$ couples to ${\cal N}\simeq 10^5$ gauge fields \cite{Anber:2009ua,Anber:2012du}. As we will see in the following sections, this result is completely modified in the presence of \textit{gravitationally-enhanced friction} term $K\neq 1$.

\subsection{Scalar perturbations}\label{sec:scalar}

\begin{figure}
	\centering
	\includegraphics[scale=0.53]{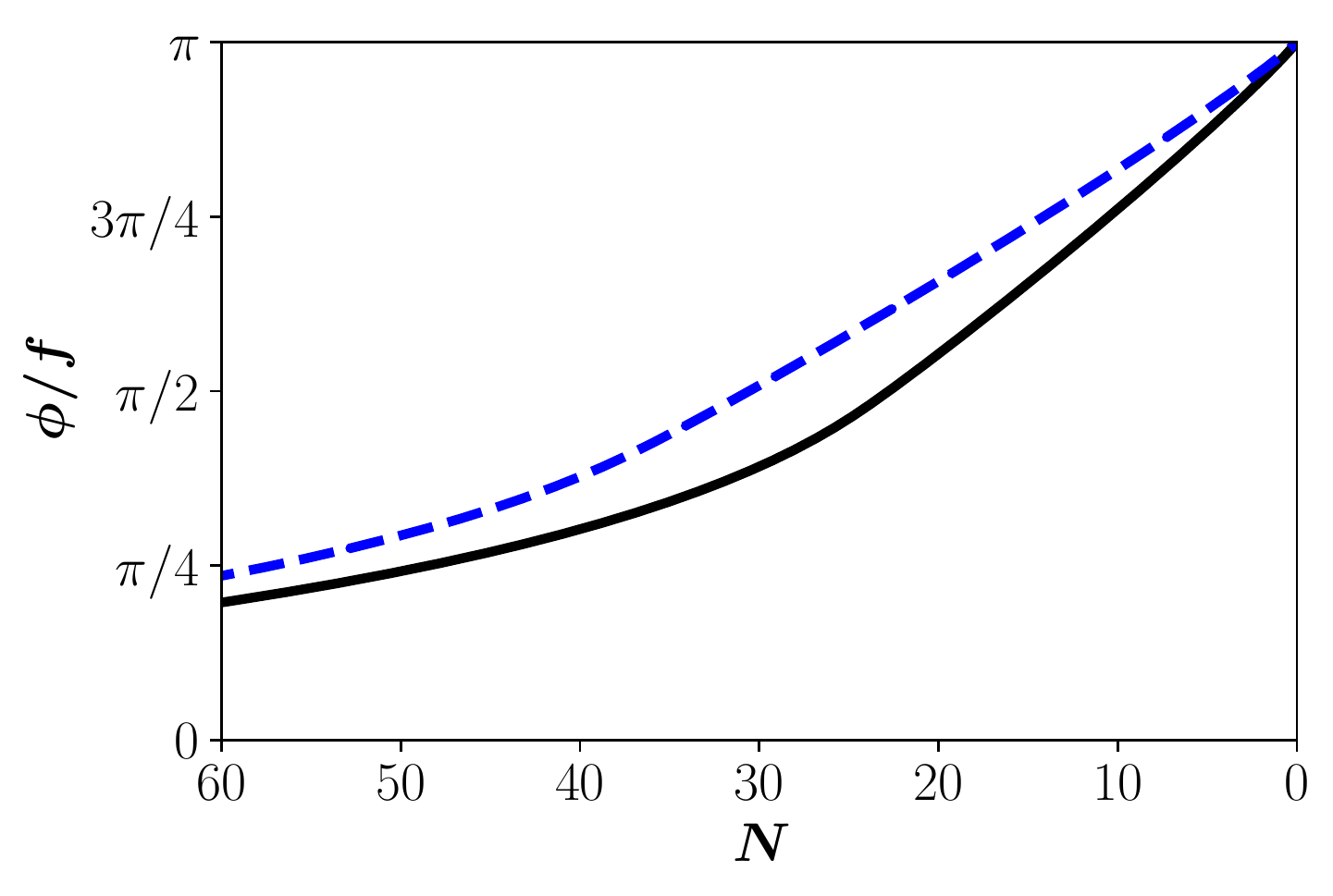}
	\includegraphics[scale=0.53]{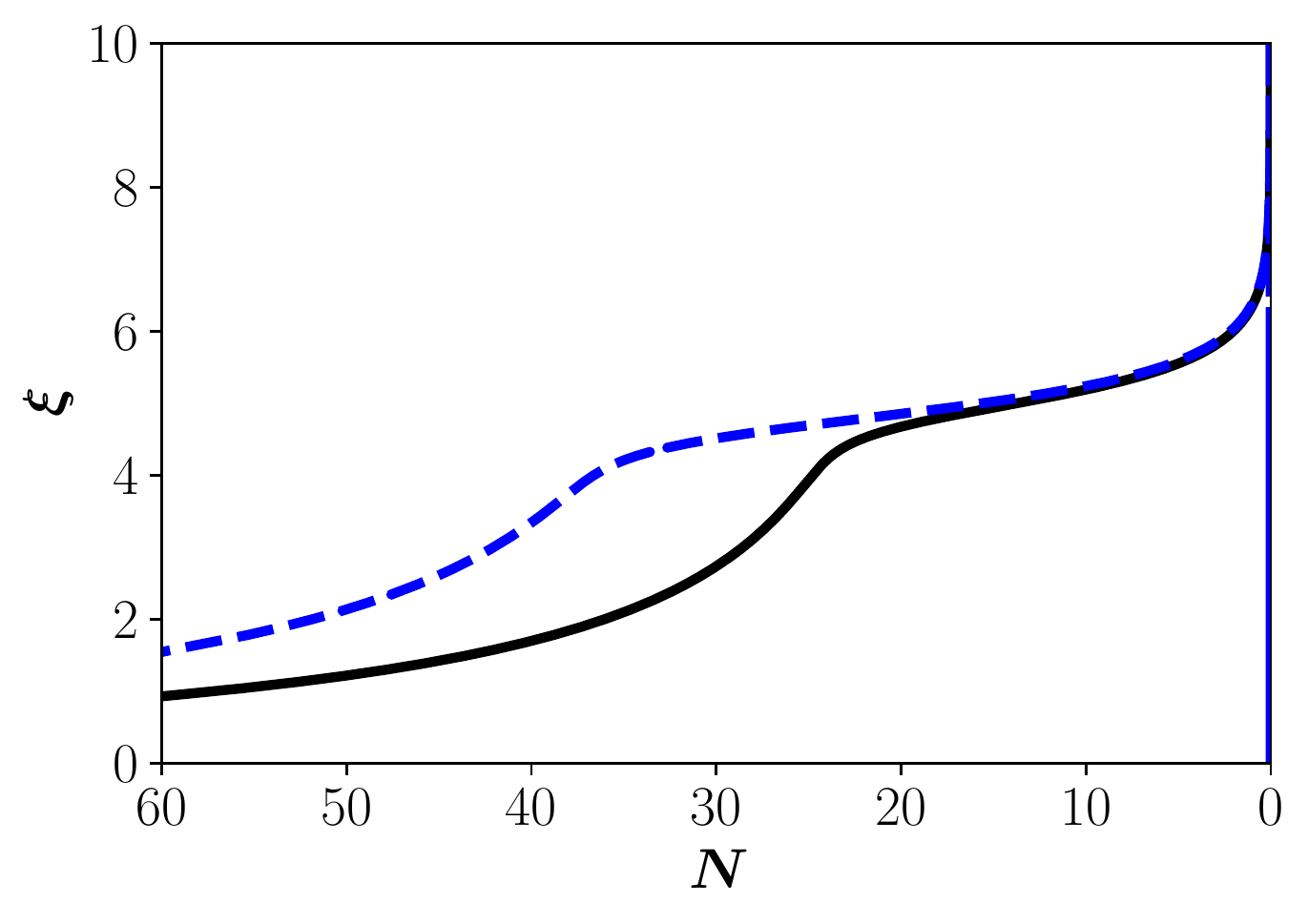}
	\includegraphics[scale=0.53]{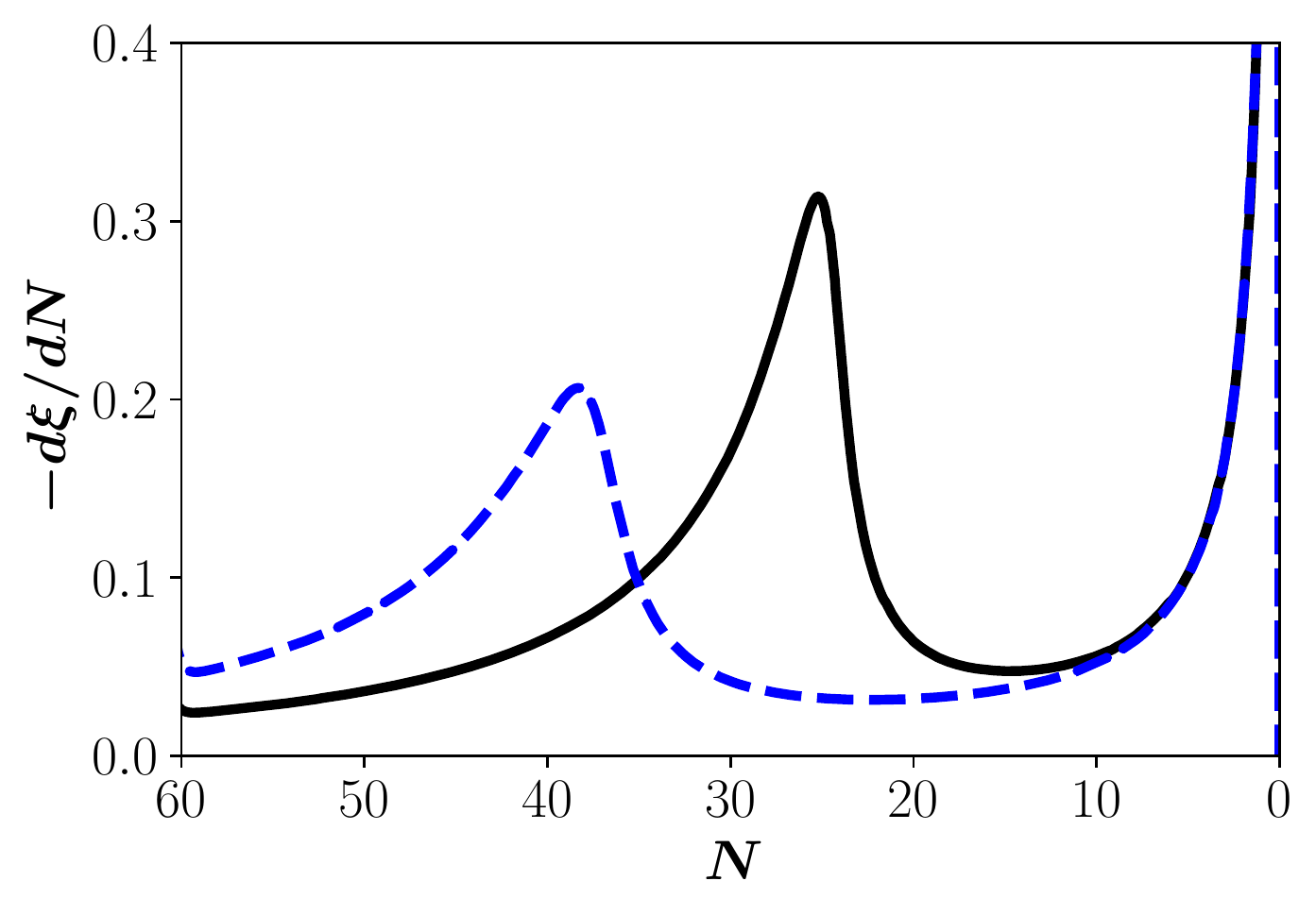}
	\includegraphics[scale=0.53]{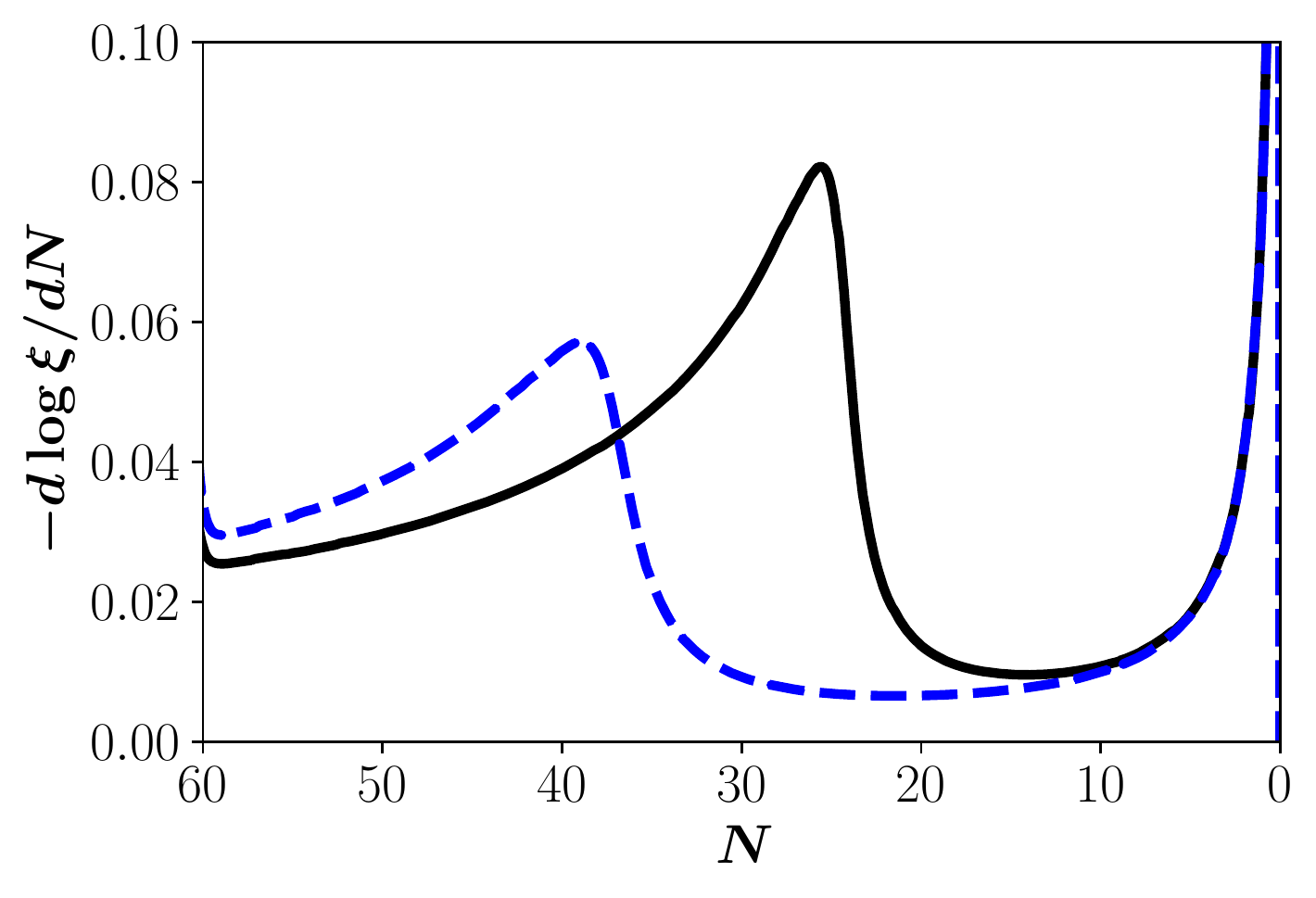}
	\caption{ {Evolution of the inflaton field $\phi$ and the instability parameter $\xi$ as a function of the number of $e$-folds $N$, for fiducial values $\Lambda=4.6\times 10^{-3}~M_P$, $f/M=1.7\times 10^5$, $\alpha=150$ (black solid lines), and $\Lambda=5.1\times 10^{-3}~M_P$, $f/M=1.5\times 10^5$, $\alpha=200$ (blue dashed lines), assuming always ${\cal N}=20$ fields. Additionally, we display the derivative and the logarithmic derivative of $\xi$.}
    }
	\label{fig:xievol}
\end{figure}

The spectrum of scalar perturbations can be straightforwardly computed in the  $\delta \phi \neq 0$  gauge~\cite{Germani:2010ux, Germani:2011ua, Ema:2015oaa} (see also Refs.~\cite{Kobayashi:2011nu, DeFelice:2011uc, Tsujikawa:2012mk}).   Following the steps in Appendix~\ref{appendix:action}, the corresponding second-order action in conformal time takes the form
\begin{equation}\label{quadphiMV}
S^{(2)}_{\delta \phi} = \int {\rm d}^3 x\, {\rm d} \tau\, \frac{1}{2}  \left[ {u'}^2 -   c_s^2\,\left(\nabla u \right)^2 +\left(  \frac{z''}{z} - m^2 a^2 \right) u^2 \right] +  \int {\rm d}^3 x\, {\rm d} \tau \,a^4\, \frac{\alpha}{f}\, \frac{u}{z}\,  \delta [ \vec{E}_a \cdot  \vec{B}_a ] \,,
\end{equation}
with
\begin{equation}\label{MSvariables}
u\equiv z\, \delta\phi\,, \hspace{20mm} z\equiv a\, F\, \sqrt{\frac{2G}{\epsilon_K}}\,,
\end{equation}
the canonical Mukhanov-Sasaki variables,
\begin{equation}\label{EFGdef}
 \epsilon_K \equiv \frac{\dot{\phi}^2}{M^2 M_P^2}\,, \hspace{10mm} F \equiv \frac{1-\frac{1}{2}\epsilon_K}{1-\frac{3}{2}\epsilon_K}\,, \hspace{10mm} G \equiv  \frac{\epsilon_K}{2} \left(1+ 3\frac{H^2}{M^2} \frac{1+\frac{3}{2}\epsilon_K}{1-\frac{1}{2}\epsilon_K} \right)\,, 
\end{equation}
and $c_s^2$ and $m^2$ effective speed and mass parameters whose explicit expressions can be found in Eqs.~\eqref{eq:cs} and \eqref{massKAxial}. The term $\delta[\vec{E}_a\cdot \vec{B}_a]$ contains two contributions associated respectively with the intrinsic inhomogeneities in $\vec{E}_a\cdot \vec{B}_a$ at $\phi =0$ and the explicit $\dot{\phi}$ dependence~\cite{Anber:2009ua, Almeida:2018pir}, namely 
\begin{equation}\label{deltadef}
\delta[\vec{E}_a\cdot \vec{B}_a] \simeq \left[\vec{E}_a\cdot \vec{B}_a - \langle \vec{E}_a\cdot \vec{B}_a \rangle  \right]_{\delta\phi =0} + \frac{\partial  \langle \vec{E}_a\cdot \vec{B}_a \rangle }{\partial \dot{\phi} } \delta  \dot{\phi} \equiv \delta_{\vec{E}_a\cdot \vec{B}_a }+ \frac{\partial  \langle \vec{E}_a\cdot \vec{B}_a \rangle }{\partial \dot{\phi} } \delta  \dot{\phi}\,.
\end{equation}
In Fourier space, the equation of motion derived from the action \eqref{quadphiMV} takes the form 
\begin{equation}\label{usourced}
u'' +   \left(c_s^2\, k^2  + m^2 a^2 - \frac{z''}{z}\right) u =  \frac{\alpha}{f} \, \frac{a^4}{z}\, \delta [ \vec{E}_a \cdot  \vec{B}_a ]\,,
\end{equation}
with $u=u(k,\tau)$  and $k$ the corresponding wave number or comoving momentum.  
This differential equation can be solved in two separated pieces: a vacuum homogeneous solution $u^{(0)}$  including the effect of the non-minimal kinetic coupling and a particular solution $u^{(\rm{s})}$ sourced by the axial contribution, i.e.  $u(k,\,\tau) = u^{(0)}(k,\,\tau)  + u^{(\rm{s})}(k,\,\tau)$. Assuming these to be statistically independent~\cite{Barnaby:2011vw}, $\left\langle u(k) u(k') \right\rangle =\langle u^{(0)}(k)  u^{(0)}(k')  \rangle +\langle  u^{(\rm{s})}(k) u^{(\rm{s})}(k') \rangle$, and neglecting order one corrections associated with the precise choice of the pivot scale, the spectrum of primordial density fluctuations can be written as (cf.~Appendix~\ref{appendix:action} for details)
\begin{equation} \label{eq:spesou} 
P_{\zeta}(k)  \simeq  \bar{P}_{\zeta} \left(-\frac{c_s k}{2aH}\right)^{\gamma_p -1}\,, 
\end{equation}
with 
\begin{equation} \label{Amplitude}
\bar{P}_{\zeta}  \simeq  \frac{\, H^4 }{4 \pi^2 K \dot{\phi}_0^2}  \left(\frac{\epsilon_K}{ 2\, G\, F^2\, c_s^3}\right)  \left[1+  \frac{4\, G\, F^2\, c_s^3{\cal F}}{ {\cal N} \epsilon_K }  \left(\frac{ \alpha\, {{\cal N}} \,  H}{ \Delta \, K  \,f }\right)^2\,  \frac{e^{4\pi \xi}}{\xi^8} \, { \left(\frac{2^6 \xi\,}{ c_s}\right)^{\gamma_p -1}}\right].
\end{equation}
Here ${\cal F}\simeq 2.13 \times 10^{-6}$ is a numerical constant coming from the momentum integral of the  $\delta_{\vec{E}\cdot\vec{B}}$ spectrum~\cite{Anber:2009ua, Almeida:2018pir},
\begin{equation} \label{nuDelta}
\gamma_p-1= 3 - \sigma \pm \Delta\,, \hspace{12mm} \sigma\equiv\frac{\pi\, \alpha\, V_{\phi} }{K f\,H^2 }\,, \hspace{12mm} 
 \Delta^2  \equiv    \left( 1- \sigma \right)^2 +   4\left(\nu^2-\frac{1}{4}-\sigma \right),
\end{equation}
and
\begin{equation}\label{nudef}
 \nu \equiv \frac{3}{2} \sqrt{1+\frac{4}{3}\epsilon_{H} +\frac{8}{9}\delta_{K} -\frac{4}{9} \frac{m^2}{H^2}}\,, \hspace{20mm}
\delta_{K} \equiv \frac92\frac{\dot{H}}{M^2 K}\,.
\end{equation} 
Note that this cumbersome expression reduces to the standard one in the decoupling limit $\alpha\to 0$, $M\to\infty$, where $F=K=c_s=1$, $G=\delta_K=\epsilon_K=0$ and $\dot{\phi} \simeq -V_{\phi}/(3H)$. For the non-vanishing $\alpha$ and finite $M$ case we are interested in, we rather have $\epsilon_K\simeq 0$, $F\simeq c_s\simeq  1$ and $G\simeq \epsilon_K\, K/2$, such that
\begin{equation}\label{totalA2}
	\bar{P}_{\zeta} \simeq \frac{H^4}{4\pi^2 \,\dot{\phi}^2 K} \left[ 1   + \frac{2 K {\cal F}}{{\cal N}}  \left(
	\frac{\, \alpha \, {\cal N}  \,  H }{ \Delta\,  K\,  f}\right)^2 \, \frac{e^{4\pi \xi}}{\xi^8}(2^6 \xi)^{\gamma_p-1} \right]\,.
\end{equation}
It is convenient to rewrite this expression as a function of the number of $e$-folds $N$ till the end of inflation. To this end, let us consider the evolution of the background inflaton field. In the strong friction [\,$H^2\simeq 2\Lambda^4/(3M_P^2)$\,] and strong axial 
[\,$H^2\simeq V/(3 M_P^2)$ and $V_{\phi} \simeq -V/f  \simeq  -\frac{ I_3\,{\cal N}\, \alpha}{f} \left(H/\xi\right)^4 e^{2 \pi \xi}$\,] regimes, the equation of motion~\eqref{eq:phi} admits an approximate solution
\begin{equation}\label{eq:phiN}
	\phi(N)\simeq
	\begin{cases}
		\phi_*\,e^{{\cal B}\,(N_*-N)} &\qquad \text{ for } \,\,\,\, N_*\geq N\gg N_c\,,\\[8pt]
		\pi\,f - \frac{2\,f\,\xi}{\alpha}N+{\cal O}(N^2) &\qquad \text{ for } \,\,\,\, N_c\gg N\geq 0\,,
	\end{cases}
\end{equation}
with  $\phi_*$ the initial field value,  $N_{*}=60$ the number of $e$-folds at which the pivot scale left the horizon,
\begin{equation}\label{Nc}
 {\cal B}	\simeq \frac14 \left(\frac{M}{f}\right)^2 \left(\frac{M_P}{\Lambda}\right)^4, \hspace{15mm}
					N_c\simeq\frac{\phi_*\,{\cal B}\,N_*+\phi_*-\pi\,f}{\phi_*\,{\cal B}-\frac{2f\,\xi}{\alpha}}\,,
\end{equation}
and 
\begin{equation}\label{eq:xi}
			\xi\simeq -\frac{2}{\pi}\,\mathcal W_{-1}\left[-\frac{\pi}{\sqrt{6}}\frac{\Lambda}{M_P}\left(\alpha\,{\cal N}\,{\cal C}\,\mathcal{I}_3\right)^{1/4}\right],
\end{equation}
with ${\cal C}$ an order one constant and $\mathcal W_{-1}(x)$  the Lambert function. The approximate solution of Eq.~\eqref{eq:phiN} fits well the result of numerically solving the system of equations~\eqref{Friedman1}, \eqref{Friedman2} and~\eqref{eq:phi}, shown in Fig.~\ref{fig:xievol} for two benchmark points. The figure displays also the evolution of $\xi$  {and its derivative}, as given by Eq.~\eqref{eq:inst_par}. This instability parameter grows large towards the end of inflation, acting as an effective friction term for the inflaton field and generating additional $e$-folds even if the potential is steep and the gravitationally-enhanced friction ceases to be important.

Taking into account the above relations and assuming $i)$ the absence of trans-Planckian masses, $f\ll M_P$, $ii)$ a weak coupling regime in the inflaton-gauge interactions, $f\gg M$, $iii)$ sub-Planckian curvatures, $H\ll M_P$, and $iv)$ a large gravitational friction at early times $H(N_*)\gg M$, the scalar power spectrum \eqref{totalA2} can be approximated as
\begin{equation}\label{eq:PN}
	\bar{P}_{\zeta}(N) \simeq
	\begin{cases}
		\frac{4}{3\pi^2}\left(\frac{f}{M}\right)^2  \left(\frac{\Lambda}{M_P}\right)^8 \left(\frac{f}{\phi}\right)^2 &\qquad\text{ for }\,\,\,\,N_*\geq N\gg N_c\,,\\[8pt]
		\frac{5\times 10^{-2}}{{\cal N}\xi^2} &\qquad\text{ for }\,\,\,\,N_c\gg N\geq 0\,,
	\end{cases}
\end{equation}
with the first and second lines corresponding respectively to the strong friction \cite{Germani:2011ua, Tsujikawa:2012mk} and the strong axial regime~\cite{Anber:2009ua,Sorbo:2011rz, Anber:2012du, Barnaby:2012xt, Almeida:2018pir}.

\subsection{Tensor power spectrum} \label{GW}

The creation of gauge fluctuations via the Chern-Simons interaction $\phi\,\tilde{F}\,F$ sources also the production of helical GW.  To determine their amount and chirality, let us consider the perturbed metric  $d {s}^2 = {a}^2({\tau})\left[- d {\tau}^2 + ( \delta_{ij} + {h}_{ij})\,  d {x}_i  d{x}_j \right]$
 with $h_{ij}$ a transverse-traceless perturbation, $h^i_i = \partial^jh_{ij} = 0$.
The variation of the quadratic expansion of the action~\eqref{KHAS} in $h_{ij}$~\cite{Germani:2011ua},
\begin{equation}\label{hquadt}
	S_{t}^{(2)} = \frac{M_P^2}{8} \int d^3 x \,d \tau \, a^2 \left[ \left( 1 - \frac{1}{2a^2} \frac{{\phi'}^2}{M^2\, M_P^2} \right) h_{ij}'^2 - \left( 1 + \frac{1}{2a^2}\frac{{\phi'}^2}{M^2\, M_P^2} \right) (\nabla {h}_{ij})^2 \right],
\end{equation}
leads to the following  equation of motion for the tensor modes in Fourier space,
\begin{equation}
h''_{ij} + \left[2 \frac{a'}{a} +  \frac{\beta'}{\beta}  \right] h'_{ij}+c^2_t  \, k^2 \, h_{ij}=\frac{2}{\beta \,M^2_P} \,T_{ij}^\text{EM}, 
\end{equation}
with
\begin{equation}\label{eq:beta}
	\beta \equiv  1- \frac{1}{2}\epsilon_{K} \,, \hspace{20mm}	c^2_{t} \equiv \frac{ 1 + \frac{1}{2}\epsilon_{K}   }{ 1- \frac{1}{2}\epsilon_{K}  }\,.
\end{equation}
Projecting into the helicity basis defined by the polarization vectors $\epsilon^i_\lambda(\vec{k})$ and using the relations 
\begin{equation}
	h^{ij}(\tau,\,\vec{k}) =\sqrt{2}\sum_{\lambda=\pm}\epsilon^i_\lambda(\vec{k})\,\epsilon^j_\lambda(\vec{k})\,h_\lambda(\tau,\,\vec{k})\,, \hspace{15mm} h_{\lambda} (\tau,\,\vec{k}) = \Pi_\lambda^{lm} h_{lm}(\tau,\,\vec{k})\,,
\end{equation}
\begin{equation}
\Pi_\lambda^{lm}=\frac{1}{\sqrt{2}}\epsilon_{-\lambda}^l(\vec{k})\epsilon_{-\lambda}^m(\vec{k})\,,
\end{equation}
the equation for the tensor modes with right- ($\lambda=+2$) and  $(\lambda=-2$) left-handed polarizations can be written as
\begin{equation}\label{eq:tenmod}
	h''_\lambda + \left[2 \frac{a'}{a} +  \frac{\beta'}{\beta}  \right]  h'_\lambda+c^2_t \, k^2 \, h_\lambda =\frac{2 }{\beta\,M^2_P}\,\Pi_\lambda^{lm}\,T_{lm}^\text{EM},
\end{equation}
with $T^\text{EM}$ the energy momentum tensor for the source part of the action~\eqref{KHAS}. Due to the projector  $\Pi_\lambda^{lm}$, only the transverse part of this quantity  contributes to the tensor perturbation, namely $T_{ij}^\text{EM, TT}= -a^2(E_i^a E_j^a + B_i^a B_j^a)$.  
In a slow-roll quasi-de Sitter regime where both $\xi$ and $H$ are approximately constant, we can rewrite Eq.~\eqref{eq:tenmod} as
\begin{equation}\label{hmodesbetaconst}
 h''_\lambda -   \frac{2}{\tau}  h'_\lambda+c_t^2\, k^2\,  \, h_\lambda =\frac{2 }{\beta\, M^2_P}\Pi_\lambda^{lm}\,T_{lm}^\text{EM}\,, 
\end{equation}
which is formally equivalent to that in Refs.~\cite{Sorbo:2011rz,Almeida:2018pir} upon the trivial replacement $1/M_P^2\to 1/(\beta M_P^2)$.  Assuming again the vacuum and sourced modes to be statistically independent~\cite{Barnaby:2011vw},  $
	\left\langle h_\lambda  (k)\, h_\lambda  (k')  \right\rangle = \langle h_\lambda^{(0)} (k) \, h_\lambda^{(0)} (k')   \rangle  + \langle  h_\lambda^{({\rm s})}(k) \, h_\lambda^{({\rm s})}(k')   \rangle$, 
the total tensor spectrum for each polarization mode can be written as $
{P}_{t,\lambda} \equiv {P}^{(0)}_{t,\lambda} +{P}^{(\rm s)}_{t,\lambda} 
$
with
\begin{equation}
\label{eq:GW_spectra}
 P^{(0)}_{t,\lambda} = \frac{k^3}{2 \pi^2} |h_{\lambda}|^2 \simeq  \frac{H^2}{\pi^2\, M_P^2  \,c_t\,  \left(1 + \frac{1}{2}\epsilon_{K} \right)}\,,  \hspace{15mm} 	P^{({}\rm s)}_{t,\lambda} \simeq   \frac{   {\cal A}_{\lambda}\,  {\cal N} }{\pi^2\, \beta^2 }\frac{H^4}{M_P^4} \frac{e^{4\pi \xi}}{\xi^6} \,,
\end{equation}
the vacuum and sourced contributions, ${\cal A}_{+}\simeq 8.6\times10^{-7}$ and ${\cal A}_{-}\simeq 1.8\times10^{-9}$ \cite{Sorbo:2011rz}. Using these expressions, we can define the tensor-to-scalar ratio $r$ and the chirality parameter or degree of polarization $\Delta \chi$, 
\begin{equation}\label{eq:r}
	r \equiv \frac{\sum_{\lambda} P_{t, \lambda} }{{P}_{\zeta}}\,, 
\hspace{20mm}
	\Delta\chi\equiv \frac{ P_{t,+}- P_{t,-}}{ P_{t,+}+ P_{t,-}}\,.
\end{equation} 
As for the scalar perturbations, we can distinguish two regimes. At early times, the friction due to the non-minimal coupling controls the dynamics, such that the contribution of the axial coupling to the tensor power spectrum is completely negligible. In this stage, the spectrum is approximately flat and non-chiral,
\begin{equation}\label{eq:GW_spectra1}
P_{t,\lambda}\simeq \frac{2}{3\pi^2}\left(\frac{\Lambda}{M_P}\right)^4\,,
\hspace{10mm}
r\simeq \left(\frac{M}{f}\right)^2  \left(\frac{M_P}{\Lambda}\right)^4 \left(\frac{\phi}{f}\right)^2\,, \hspace{10mm} \Delta\chi \simeq 0\,.
\end{equation}
At later times, the exponential amplification of the $A_+$ modes as the field velocity increases makes them the dominant source of tensor perturbations~\cite{Sorbo:2011rz, Anber:2012du, Barnaby:2012xt, Almeida:2018pir}, providing an enhanced parity-violating GW background with 
\begin{equation}\label{eq:GW_spectra2}
P_{t,\lambda}\simeq \frac{4\mathcal A_\lambda\mathcal N}{9\pi^2 \beta^2}\left(\frac{\Lambda}{M_P}\right)^8  { \frac{e^{4\pi \xi}}{\xi^6} \simeq  \frac{9\mathcal A_\lambda}{\pi^2 \mathcal N I_3^2 \alpha^2 \beta^2} \xi^2},\,  \hspace{5mm}
r\simeq 2.9\times 10^2 \, \frac{\xi^4}{\alpha^2}  \,,  \hspace{5mm}     \Delta\chi\simeq\frac{\delta\chi}{1+\delta\chi}\,,
\end{equation}
where we have again assumed  {the strong axial regime approximation to be valid [\,$H^2\simeq V/(3 M_P^2)$ and $V_{\phi} \simeq -V/f  \simeq  -\frac{ I_3\,{\cal N}\, \alpha}{f} \left(H/\xi\right)^4 e^{2 \pi \xi}$\,]}  
and defined 
\begin{equation}
\delta\chi\equiv \frac{{\cal N}{\cal A}_+}{3}\left(\frac{\Lambda}{M_P}\right)^4\frac{e^{4\pi\,\xi}}{\xi^6}  { \simeq   \frac{27 {\cal A}_{+} M_p^4 \xi^2 }{2I_{3}^2 {\cal N} \alpha^2 \beta^2} }\,.
\end{equation}

\section{Phenomenology}\label{sec:parameter}

In order to assess the testability of our scenario and determine the viable parameter space, we confront now the scalar and tensor power spectra obtained in Section \ref{sec:spectra} with present and future data sets. On the one hand, we will enforce the compatibility of our predictions with current CMB observations. On the other hand, we will consider small-scale restrictions coming from PBH formation. Finally, we discuss the potential detection of chiral GW by future GW experiments. 

\subsection{Cosmic microwave background}

The precise measurements of CMB anisotropies \cite{Akrami:2018odb} impose strong constraints on the primordial power spectrum at scales $0.008~{\rm Mpc}^{-1} \lesssim k \lesssim 0.1~{\rm Mpc}^{-1}$, providing exquisite information on the first 7 $e$-folds of inflation. This range of knowledge is extended to about 20 $e$-folds by measurements of Lyman-$\alpha$ forest and $\mu$ spectral distortions,\footnote{These are related to the energy injection into the photon-baryon plasma from primordial perturbations reentering the horizon at redshift $2 \times 10^6 \lesssim z \lesssim 5 \times 10^4$.} which are sensitive to the integrated scalar power spectrum in the range $50~{\rm Mpc}^{-1} \lesssim k \lesssim 10^4~{\rm Mpc}^{-1}$ \cite{Hu:1994bz}.

\begin{figure}
	\centering
	\includegraphics[scale=0.49]{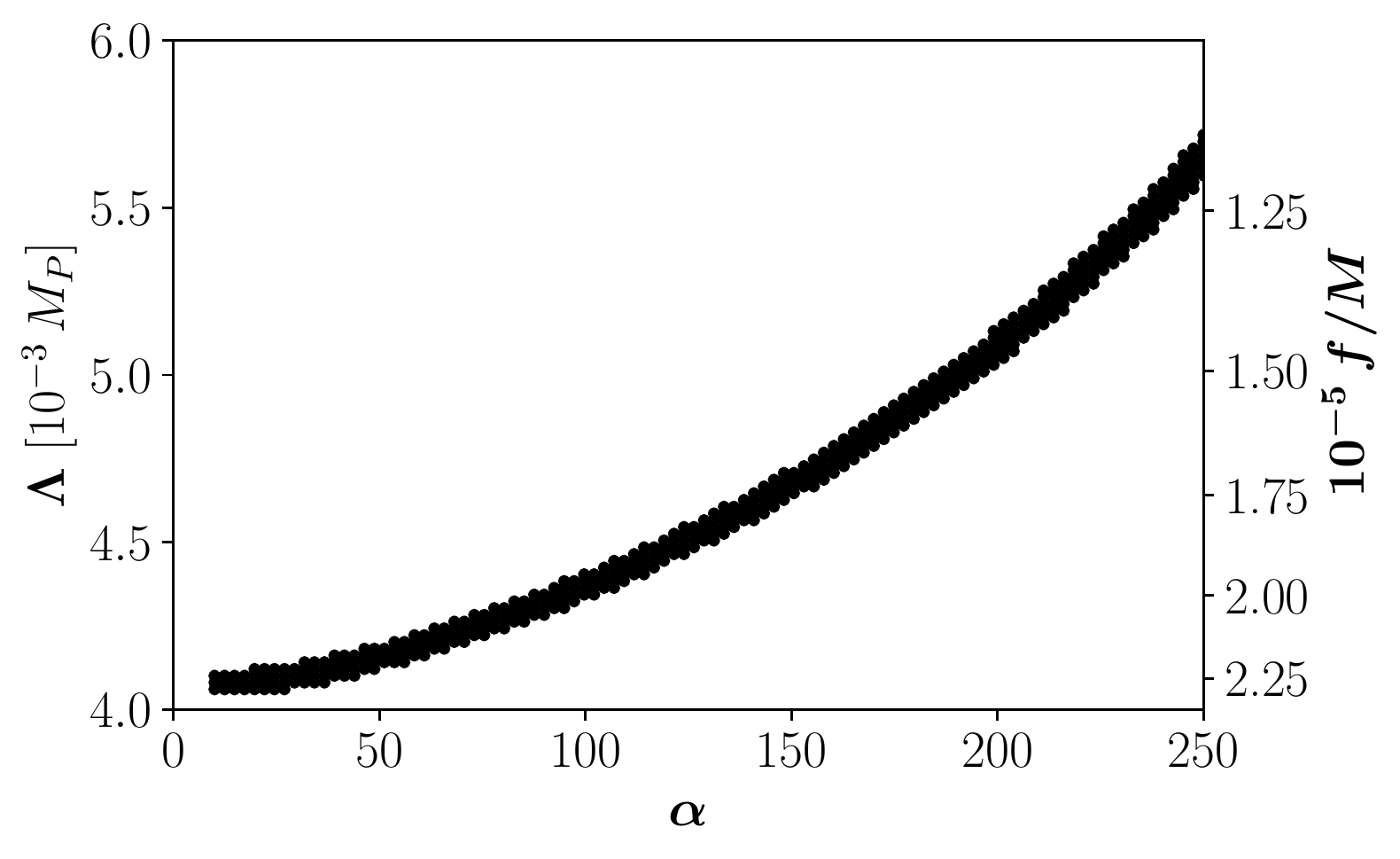}
	\includegraphics[scale=0.49]{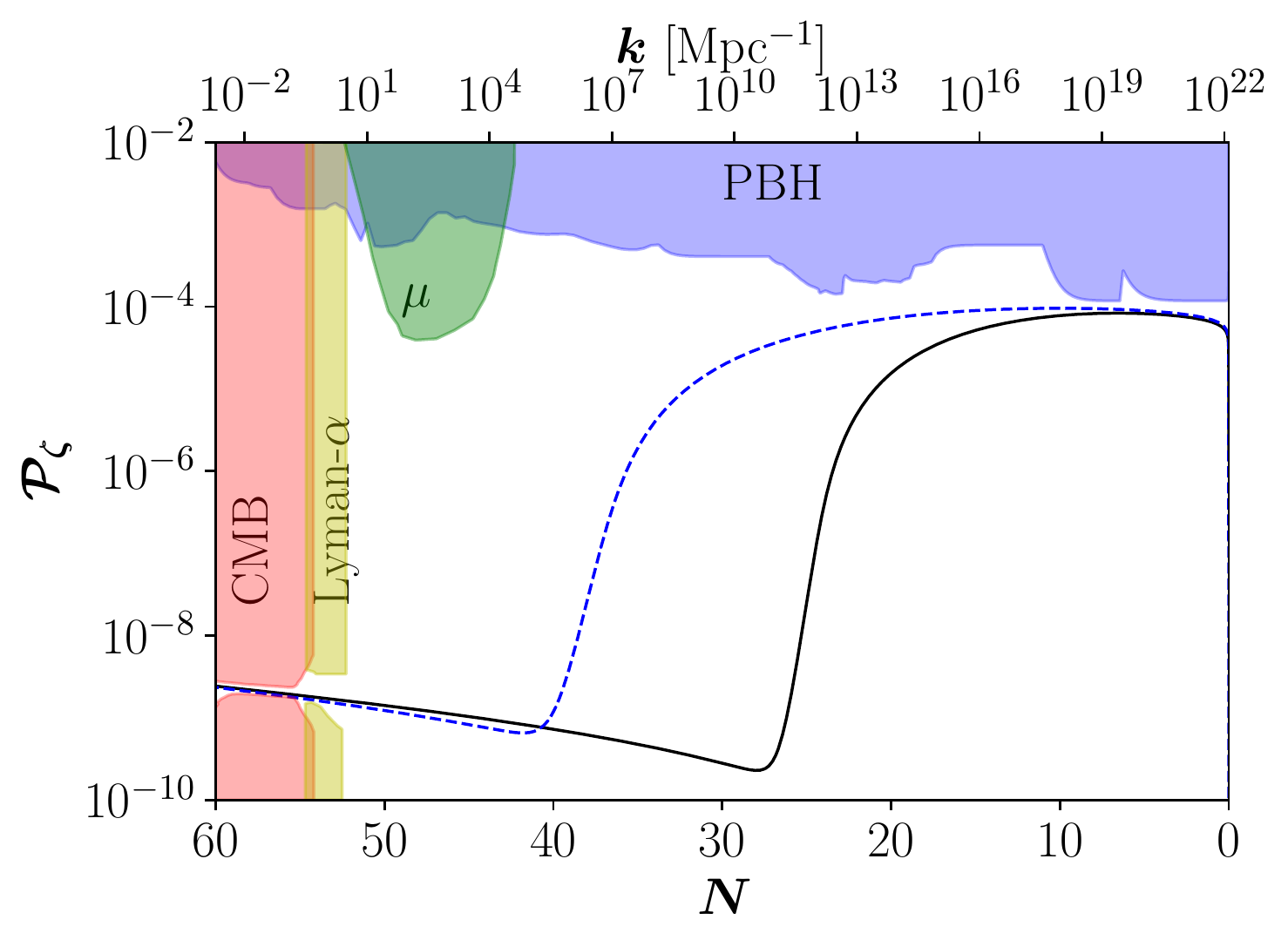}
	\caption{Left panel: Parameter space compatible with Planck results on the amplitude and tilt of primordial density fluctuations, assuming $N_*=60$ and ${\cal N}=20$.	Right panel: Scalar spectra for the benchmark points used in Fig.~\ref{fig:xievol}.
	The colored regions are excluded by CMB anisotropies (red) \cite{Akrami:2018odb}, Lyman-$\alpha$ forest (yellow) \cite{Bird:2010mp}, $\mu$ distortions (green) \cite{Kohri:2014lza, Carr:2020gox} and PBH bounds (blue) \cite{Carr:2020gox, Allahverdi:2020bys}.
	}
	\label{fig:numevolP}
\end{figure}

The parameter space compatible with the latest Planck results on the amplitude and tilt of primordial density fluctuations  \cite{Akrami:2018odb} is displayed in the left panel of Fig.~\ref{fig:numevolP} for $N_*=60$ and ${\cal N}=20$ (cf.~Section \ref{sec:PBH}).  Note that the allowed region could be additionally constrained by the kinematic restriction \eqref{OmegaK} in Appendix \ref{app:slowroll}. For the ${\cal O}(1)$ values of $\xi$ in this regime, this restriction is very mild, $\alpha\gtrsim 7.5$. 

For illustration purposes, we display also exemplary power spectra in the right panel of Fig.~\ref{fig:numevolP}. The initial values $\phi_*$ and $\dot\phi_*$ for the inflaton field and its derivative were fixed using its equation of motion~\eqref{eq:phi} and the COBE normalization.  Additionally, the parameters $\Lambda$ and $f/M$ are related by the spectral tilt 
\begin{equation}\label{eq:ns}
	n_s-1 \equiv \frac{d \ln P_\zeta (k)}{d\ln k} \simeq -2\,{\cal B}
\end{equation}
around $N_*=60$. In this region, the power spectrum is fairly Gaussian since the dominant gravitationally-enhanced friction in Eq.~\eqref{eq:phi} could be easily transferred to the axion potential~\eqref{eq:pot} by performing a suitable field redefinition~\cite{Germani:2011ua}. This reduces the scenario to the standard one with no extra scales and slow-roll suppressed non-Gaussianities \cite{Maldacena:2002vr}. Note that this argument does not apply to the upward bent of the spectrum at small scales, where non-Gaussianity plays indeed a very important role \cite{Cook:2013xea}, as we now proceed to discuss. 

\subsection{Primordial black holes}\label{sec:PBH}

The axial enhancement of the primordial power spectrum at sub-CMB scales (cf. right panel of  Fig.~\ref{fig:numevolP}) can lead to the formation of PBH with masses of the order of the horizon mass at the time of reentry \cite{Linde:2012bt, Garcia-Bellido:2016dkw, Domcke:2017fix}. Assuming radiation domination to start immediately after the end of inflation and disregarding any mass growth due to merging or accretion, we have~\cite{Carr:2009jm, Linde:2012bt, Garcia-Bellido:2016dkw} 
\begin{equation}\label{eq:PBHmass}
M_\text{PBH}(N)\simeq 4 \pi \gamma \, M_P \, \left(\frac{M_P}{H}\right)\left( \frac{H_{\rm end}}{H}\right)\, e^{2 N}\simeq  55 \, \textrm{g} \, \gamma \, \left(\frac{10^{-6}   M_P}{H}\right)\left( \frac{H_{\rm end}}{H}\right)\, e^{2 N},
\end{equation}
with $H=H(N)$ and $\gamma \simeq 0.4$ a parameter encoding the efficiency of the gravitational collapse~\cite{Green:2004wb, Carr:2016drx}.  While PBH with masses below $10^{11}$~g decay before big bang nucleosynthesis, heavier black holes can play the role of dark matter\footnote{ {For a general and comprehensive review see e.g. Ref.~\cite{Carr:2020xqk}. For a summary} of new space- and ground-borne electromagnetic instruments potentially able to test this appealing hypothesis within the next decade, see, for instance, Ref.~\cite{Ali-Haimoud:2019khd}.} and are severely constrained by direct searches \cite{Carr:2020gox,Carr:2020xqk}. These bounds restrict the primordial power spectrum on scales much smaller than those currently probed by CMB and large scale structure surveys, providing an invaluable information on the last 40 $e$-folds of inflation. 

The formation of a PBH is a rare event. For a given perturbation amplitude in the scalar power spectrum, the fraction of causal regions collapsing into PBH is given by 
\begin{equation}\label{eq:betaM}
\beta  =   \int_{\zeta_c}^\infty {\cal P} \left( \zeta_k \right) \, d \zeta_k \,,
\end{equation} 
with $ {\cal P} \left( \zeta_k \right)$ the probability density of perturbations and $\zeta_c\sim 0.5$ a critical threshold~\cite{Green:2004wb, Garcia-Bellido:2017mdw}, within the range $\zeta_c\sim {\cal O}(0.05-1)$   \cite{Garcia-Bellido:2017aan,Garcia-Bellido:2017mdw,Domcke:2017fix} suggested by the analytic and numerical studies in Refs.~\cite{Carr:1975qj,Carr:2016drx}. As shown in Section \ref{sec:scalar}, the enhanced part of the power spectrum \eqref{totalA2} can be well-approximated by its axial contribution~\eqref{eq:PN}, which is generated by the convolution of two Gaussian modes ($A+A\to \delta\phi$) and obeys therefore a $\chi^2$ statistics \cite{Linde:2012bt}.  For this distribution, Eq.~\eqref{eq:betaM} becomes
\begin{equation}\label{betachi2}
\beta_{\chi^2} (N)={\rm Erfc} \left( \sqrt{\frac{1}{2}+\frac{\zeta_c}{\sqrt{2P_{\zeta}(N)} } } \right) \;, 
\end{equation}
with  ${\rm Erfc } \left( x \right) \equiv 1 - {\rm Erf } \left( x \right)$ the complementary error function. 
Together with Eq.~\eqref{eq:PBHmass}, this expression allows to convert the primordial power spectrum $\bar P_\zeta$ into a limit on the PBH abundance and vice versa. We follow here the second approach and display in the right panel of Fig.~\ref{fig:numevolP} the restrictions on the power spectrum following from present PBH bounds~\cite{Carr:2020gox}. The minimal number of fields needed to pass these constraints turns out to very moderate (${\cal N}\simeq 20$) and can be easily accommodated in usual grand unified groups such as $SU(N)$ without significantly altering the treatment presented here.\footnote{Note that the non-Abelian character of gauge interactions does not play a significant role since self-interactions are higher order and can be consistently neglected for weak gauge couplings.}

The above analysis is just meant to illustrate the versatility of our scenario and should be not be understood as a precise one. For the sake of clarity, there are several subtleties that should be mentioned here: 
\begin{enumerate}
    \item \textit{Slow-roll validity}. It is a well-established fact that the production of PBHs in single-field scenarios driven by a potential requires the existence of an inflection point along the inflationary trajectory where the slow-roll conditions are severely violated \cite{Motohashi:2017kbs, Germani:2017bcs}. We emphasize that this important result, usually stated as a no-go theorem, does not directly apply to the problem at hand. In particular, the enhancement of the primordial power spectrum at small scales is \textit{not} related to any special feature of the potential but rather to the explosive production of gauge fluctuations which subsequently source curvature perturbations. The validity of the slow-roll approximation during the whole inflationary evolution is illustrated in the lower panel of Fig.~\ref{fig:slow}, where we display the result of numerically solving the system of equations~\eqref{Friedman1}, \eqref{Friedman2} and~\eqref{eq:phi}.
\begin{figure}
	\centering
	\includegraphics[scale=0.53]{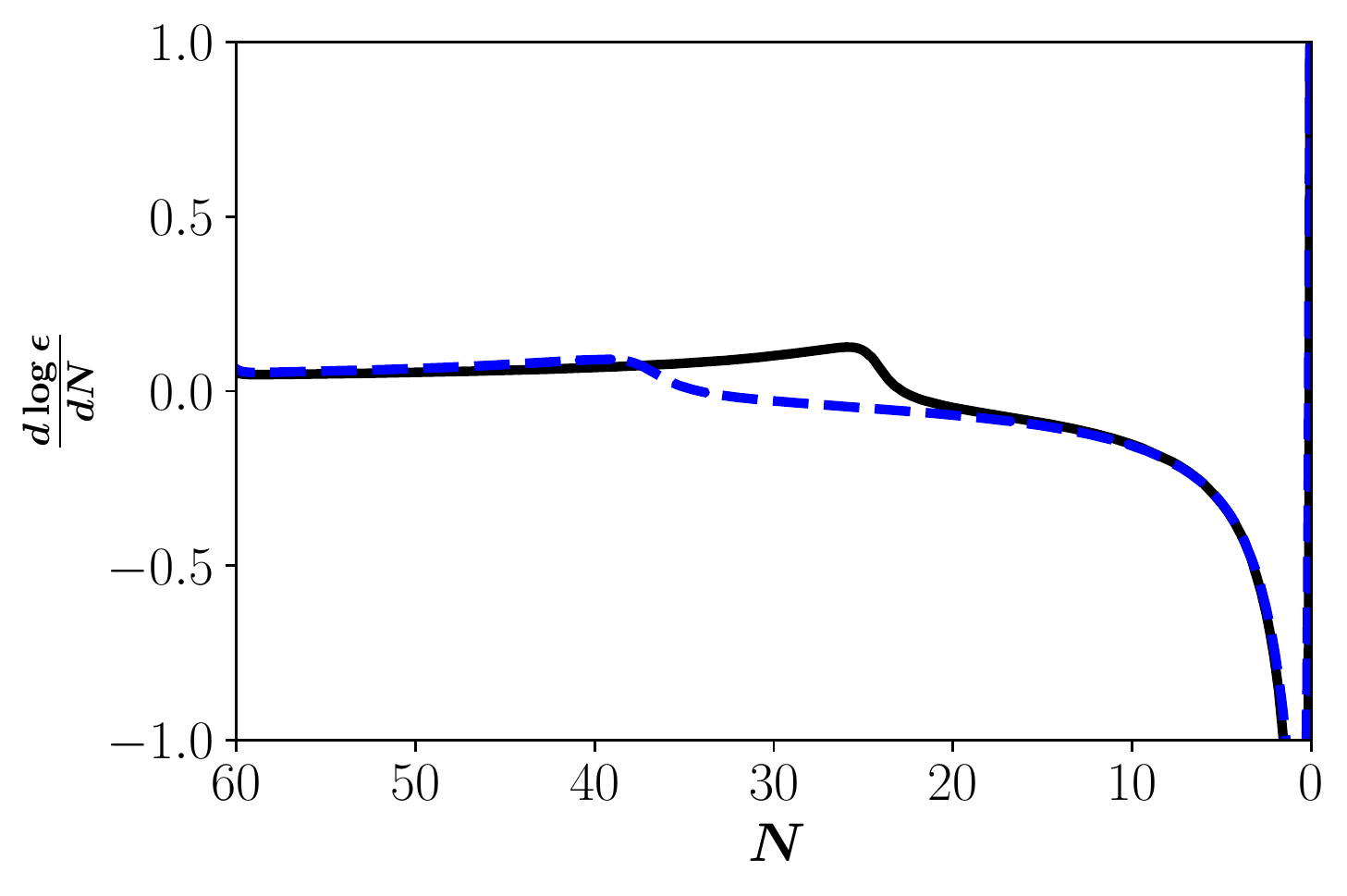}
	\caption{ {Evolution of the derivative of the slow-roll parameter $\epsilon=-\dot H/H^2$ as a function of the number of $e$-folds $N$, for the benchmark points used in Fig.~\ref{fig:xievol}.}
    }
	\label{fig:slow}
\end{figure} 
    \item \textit{Critical threshold}. We followed closely the approach of Refs.~\cite{Garcia-Bellido:2017aan, Garcia-Bellido:2017mdw, Domcke:2017fix}, employing a critical threshold $\zeta_c\sim 0.5$. A precise calculation  should take into account, however, the full shape of the  correlation functions since $\zeta_c$ depends strongly on the real space distribution of overdensities \cite{Musco:2018rwt, Escriva:2019nsa, Germani:2018jgr, Germani:2019zez}, being lower for broader shapes where pressure gradients play are less important role.
    \item \textit{Non-Gaussianities}. The formation of PBHs is inevitably a non-Gaussian process \cite{Atal:2018neu, DeLuca:2019qsy} and the precise calculation of the abundance should involve a priori the convolution of two non-Gaussian modes instead of two Gaussian ones.
    \item  \textit{Post-inflationary evolution}. We are implicitly assuming a standard $\Lambda$CDM scenario where gravity is the only attractive force for the clumping of matter. Alternative cosmologies involving fifth forces and/or deviations from radiation domination during the earliest post-inflationary epochs \cite{Allahverdi:2020bys} could affect the above estimates, allowing even for the formation of PBHs without the need of a significant enhancement of the primordial power spectrum of density fluctuations \cite{Amendola:2017xhl, Bonometto:2018dmx, Savastano:2019zpr}.
\end{enumerate}
A precise and self-consistent computation should take into account the aforementioned details while involving a combination of numerical and analytical approaches, as done for instance in Refs.~\cite{Escriva:2019nsa, Escriva:2019phb, Escriva:2020tak} for standard $\Lambda$CDM scenarios. However, it is not in the scope of this work to go beyond the approximation implied by the $\chi^2$ statistics \eqref{betachi2} \cite{Garcia-Bellido:2017aan, Garcia-Bellido:2017mdw} and the choice of a particular threshold $\zeta_c$, since the potential changes on the restrictions on the shape and amplitude of the primordial power spectrum are expected to be degenerate with the model parameters controlling the transition time $N_c$ in Eq.~\eqref{Nc} and the number of fields ${\cal N}$, respectively. A more complete and detailed discussion is left for a future research work.

\subsection{Gravitational waves}

Gravitational waves are probably the most promising relic to probe the unknown early Universe. Interestingly, the late-time amplification of the inflationary power-spectrum in the strong axial regime opens the possibility of obtaining an observable chiral GW signal in the frequency range probed by future terrestrial and space interferometers.  The associated fractional energy density per logarithmic frequency interval $f_{\rm GW}  = k / 2 \pi$ is given by~\cite{Barnaby:2011qe} 
\begin{equation}\label{OmegaGW}
\Omega_{\rm GW} \equiv \frac{1}{\rho_c} \, \frac{\partial \rho_{\rm GW,0}}{\partial \log \, k} = \frac{\Omega_{\rm R,0}}{24} \sum_\lambda P_{t,\lambda} \,,
\end{equation} 
with $\Omega_{{\rm R},0} \, h^2 \simeq 4.2 \times 10^{-5}$ the radiation density parameter today,  $P_{t,\lambda}$ the GW power spectra in Eq.~\eqref{eq:GW_spectra} and 
\begin{equation}
N  = N_*  - 44.92 + {\rm ln } \left( \frac{k_*}{0.002 \, {\rm Mpc}^{-1}} \right) - {\rm ln } \left( \frac{f_{\rm GW} }{100 \, {\rm Hz}} \right) + 
{\rm ln } \left( \frac{H_N}{H_*} \right) \,,
\label{N-f}
\end{equation} 
with $N_*=60$ and $H_N$ the value of the Hubble rate $N$ $e$-folds before the end of inflation. 

\begin{figure}
	\centering
	\includegraphics[scale=0.53]{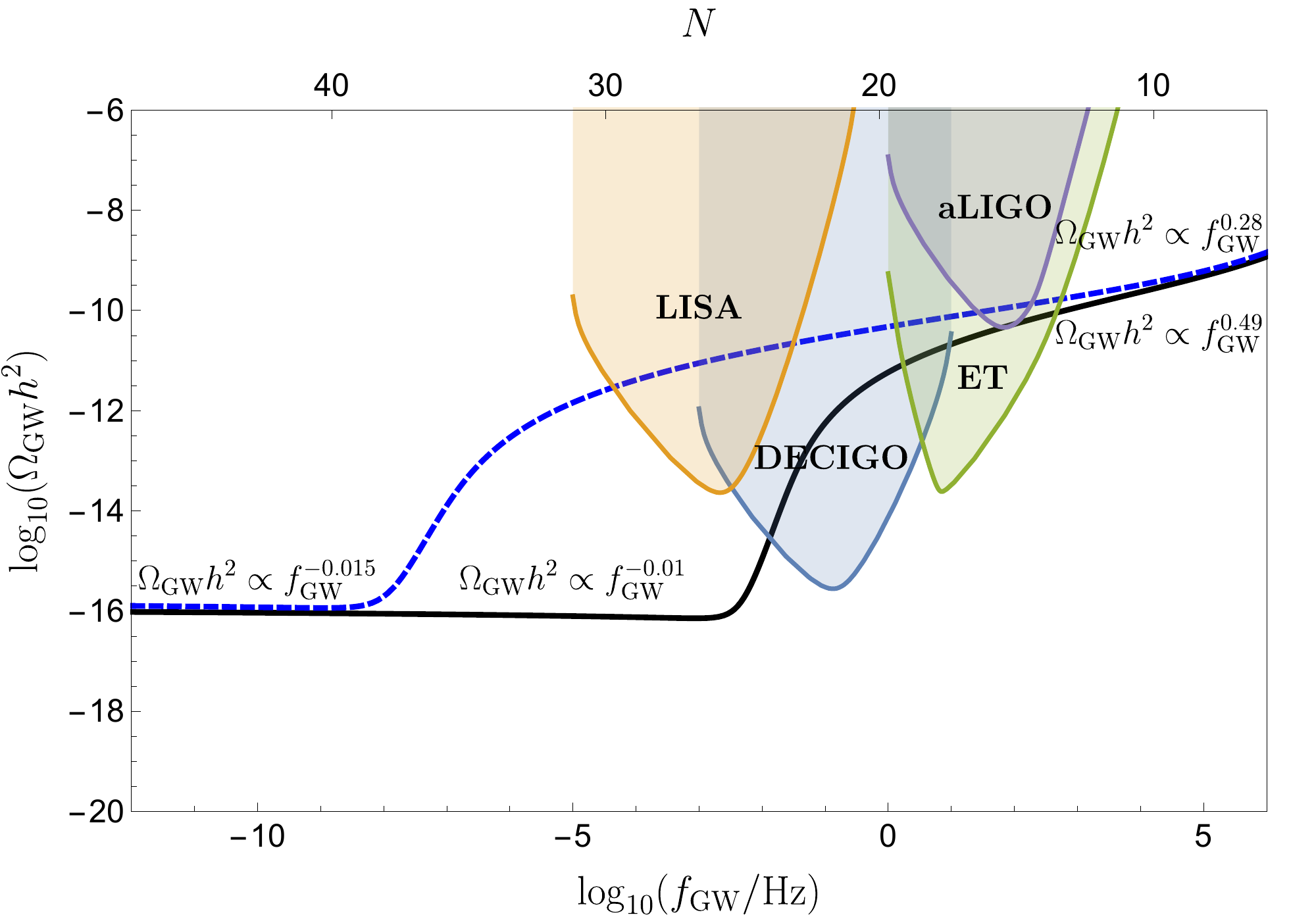}
	\caption{GW spectra for the benchmark points used in Fig.~\ref{fig:xievol}.
	The asymptotic power law behavior is obtained from the analytical approximations in Eq.~\eqref{eq:nt}.
	The experimental power-law integrated sensitivity curves account for the enhancement in detector sensitivity following the integration over frequency on top of the integration over time \cite{Thrane:2013oya, Breitbach:2018ddu}.
    \label{fig:GW}}
\end{figure}

The frequency dependence in Eq.~\eqref{OmegaGW} takes the schematic form $\Omega_{\rm GW}h^2\simeq f_{\rm GW}^{n_t}$, with 
\begin{equation}\label{eq:nt}
    n_t(f_{\rm GW} )=\frac{d\ln\Omega_{\rm GW}h^2}{d\ln f_{\rm GW} }
\end{equation}
the tensor spectral tilt, understood as a time-dependent quantity. During the first stages of the evolution $(N_*\geq N\gg N_c)$ the tensor power spectrum is dominated by the vacuum contribution in Eq.~\eqref{eq:GW_spectra}. In this regime, and at the leading order in the slow-roll parameters $\epsilon_H = -\dot H/H^2$ and $\eta\equiv -\ddot \phi/(H\dot\phi)$, we have
\begin{equation}\label{eq:epsilon}
    n_t(f_{\rm GW})\simeq -2\epsilon_H =- \left(\frac{M}{f}\right)^2\left(\frac{M_P}{\Lambda}\right)^4\,,
\end{equation}
up to mild corrections associated with the field evolution displayed in  Fig.~\ref{fig:xievol}. On the other hand, the tensor spectral tilt at late times ($N\ll N_c$) is rather given by \cite{Bartolo:2016ami},
\begin{equation}\label{eq:ntaxial}
    n_t (f_{\rm GW})   =-4\epsilon_H +4\pi\frac{d \xi}{dN}-6\frac{d\ln\xi}{dN}\,.
\end{equation}
Since the backreaction of the gauge fields on the Hubble rate is small almost till the end of inflation (cf. Appendix \ref{app:slowroll}), we can still approximate the slow-roll parameter $\epsilon_H$ by its value in Eq.~\eqref{eq:epsilon} during this regime, i.e. assume inflation to be still driven by the potential. On the other hand, from Fig.~\ref{fig:xievol} we can infer that a good approximation to the $\xi$ evolution during this stage is $\xi = a N +b$. In the absence of a simple analytical solution, these coefficients have to be extracted by fitting the numerical solution for $\xi$. However, as can be seen from Fig.~\ref{fig:GW}, the power-law approximation provides a reasonably good fit to the slope of the GW spectrum. 

The GW spectra generated by our mechanism for the benchmark points in Figs.~\ref{fig:xievol} and~\ref{fig:numevolP} are compared in Fig.~\ref{fig:GW} with the power-law integrated sensitivity curves\footnote{These curves account for the enhancement in detector sensitivity following the integration over frequencies on top of the integration over time.} of different experiments. The 
Laser Interferometer Space Antenna (LISA)~\cite{Audley:2017drz} is mostly sensitive to frequencies $10^{-4}\lesssim f\lesssim 10^{-1}$ Hz, corresponding to momenta $10^{11} \, {\rm Mpc}^{-1} \lesssim k\lesssim 10^{14}\, {\rm Mpc}^{-1}$ or an $e$-fold range $22 \lesssim N\lesssim 28$. The Deci-hertz Interferometer Gravitational wave Observatory (DECIGO)~\cite{Seto:2001qf, Kawamura:2020pcg}, the advanced LIGO detector (aLIGO)~\cite{Harry:2010zz} and the Einstein Telescope (ET)~\cite{Sathyaprakash:2012jk} extend this range all the way up to $\sim 10^2-10^3$ Hz, corresponding to scales around $k\sim 10^{17}-10^{18} \, {\rm Mpc}^{-1}$ and $N\sim 15$. Note that, although our GW spectra display generically a ``knee'' rather than a peak structure, they are maximally chiral and non-Gaussian, which could serve as a smoking gun for this mechanism and facilitate their discrimination against astrophysical backgrounds (see e.g. Refs.~\cite{Seto:2007tn, Crowder:2012ik, Thrane:2013kb, Smith:2016jqs}).  Although the measurement of the polarization is more challenging than the measure of the total spectrum, there are good prospects of detection with future experiments such as LISA or ET for sufficiently large amplitude of the total spectrum \cite{Domcke:2019zls,Ellis:2020uid}.\\

Overall, the synergy of shift-symmetric interactions advocated in this paper reconciles Natural Inflation scenarios with well-motivated UV expectations, while providing a rich phenomenology over a humongous range of scales. The natural enhancement of scalar and tensor perturbations taking place at the end of inflation translates into the simultaneous production of a dark matter component in the form of PBH and a chiral GW signal within the reach of future GW interferometers, opening the possibility of testing the model with sub-CMB physics. Our estimates rely, of course, on the accuracy of Eqs.~\eqref{eq:EB} and \eqref{usourced}, meaning that ${\cal O}(1)$ corrections should be certainly expected, especially in the large $\xi$ regime. A fully numerical computation along the lines of Refs.~\cite{Adshead:2015pva, Cheng:2015oqa, Adshead:2016iae, Notari:2016npn, Figueroa:2017qmv, Cuissa:2018oiw} will most likely be required to obtain precise results. Having presented the main qualitative features of our scenario, we postpone this detailed study to a future publication.

\section*{Acknowledgments}
We thank Juan García-Bellido  and Cristiano Germani for useful comments and discussions. NB is partially supported by the Spanish MINECO under Grant FPA2017-84543-P.
DB acknowledges support from the \textit{Atracci\'on del Talento Cient\'ifico} en Salamanca programme and from project PGC2018-096038-B-I00 by Spanish Ministerio de Ciencia, Innovaci\'on y Universidades.
This work was partly supported by COLCIENCIAS-DAAD grant 110278258747 RC774-2017 and by Universidad Antonio Nari\~no grant numbers 2018204, 2019101 and 2019248
and the MinCiencias grant 80740-465-2020.

\appendix 

\section{Action for scalar perturbations}\label{appendix:action}
In this Appendix, we present a detailed derivation of the total spectrum of primordial density fluctuations in Eqs.~\eqref{eq:spesou} and \eqref{Amplitude}. Following the standard Arnowitt-Deser-Misner (ADM) approach \cite{Baumann:2018muz}, we consider the metric decomposition 
\begin{equation}
{\rm d}s^2 =- N^2\,{\rm d}t^2 +\gamma_{ij}\,\left({\rm d}x^i + \beta^{i} {\rm d}t\right)\,\left({\rm d}x^j + \beta^{j} {\rm d}t\right),
\end{equation}
together with the gauge choice $N= 1+ n\,;  \beta_{i} = \partial_{i}\psi\,;   \gamma_{ij} = a^2(t)\,\delta_{ij}\,;  \phi (x^i,\,t)= \phi_{0}(t) + \delta \phi (x^i,\,t)\,, $
with $\phi_{0} (t)$ the homogeneous background component of the inflaton field. After expanding up to second order and integrating by parts, the scalar part of the action~\eqref{KHAS} is rewritten as \cite{Ema:2015oaa}
\begin{equation}\label{quadphi}
S^{(2)}_{\delta \phi} = \int {\rm d}^4 x\, a^3 \frac{F^2\,G}{\epsilon_{K}} \left[ \delta \dot{\phi}^2 - \frac{c_s^2}{a^2}\, (\nabla \delta \phi)^2 - m^2\, \delta \phi^2 \right] +  \int {\rm d}^4 x\, a^3\, \frac{\alpha}{f}\,   \delta[\vec{E}_a \cdot  \vec{B}_a ] \delta \phi\,.
\end{equation}
Here the equations for $\psi$ and $n$ are solved in terms of $\delta \phi$ and plugged back in the quadratic action. By doing that, one obtains the effective speed of sound $c_s$ and mass $m$, namely
\begin{equation}\label{eq:cs}
 c_s^2 =  \frac{\epsilon_K}{2 \left(1-\frac{1}{2}\epsilon_K\right)G} \left[  \left( 1+ \frac{3}{2}\epsilon_K \right) K  +    \frac{2H^2 \epsilon_K}{M^2 F}  + 6 \frac{\dot{H}}{M^2} \left( 1-\frac{1}{2}\epsilon_K \right) \right]\,, 
\end{equation}
\begin{eqnarray} \label{massKAxial}
m^2 &\equiv& \frac{\epsilon_K}{2 G\, F^2} \left[ V_{\phi\phi} \,F +\frac{ V_{\phi}  }{ M_P } \frac{M}{H} \frac{\sqrt{\epsilon_K}}{1-\frac{3}{2}\epsilon_K} \left( K+ 3\frac{H^2}{M^2} + \frac{ 1+\frac{3}{2} \epsilon_K }{ 1-\frac{3}{2} \epsilon_K } \left( \frac{1}{2}K +  \dot{H}\frac{ 1-\frac{1}{2} \epsilon_K }{ M^2 \epsilon_K } \right) \right)  \right. \\ \nonumber
	&&-  \left. \frac{ K^2  M^2 \epsilon_K ( M^2 \epsilon_K + 6 H^2 (3\epsilon_K -1)) }{ 4 H^2 ( 1-\frac{3}{2} \epsilon_K )^2 }  - \frac{1}{a^3} \frac{{\rm d}}{ {\rm d}t } \left( \frac{a^3 K M^2 \epsilon_K \left( K + \frac{\epsilon_K}{2} \left( \frac{9H^2}{M^2} -1\right) \right)}{2H (1-\frac{3}{2} \epsilon_K)^2} \right)\right],
\end{eqnarray} 
where $K$, $\epsilon_K$, $F$ and $G$  are defined in Eqs.~\eqref{Kdef} and \eqref{EFGdef}, and $\delta[\vec{E}_a \cdot  \vec{B}_a]$ includes  the first-order perturbations of the axial term $\phi\,F\,\tilde{F}$, cf.~Eq.~\eqref{deltadef}.  Introducing the canonical Mukhanov-Sasaki variables~\eqref{MSvariables}, integrating by parts and performing some algebraic manipulations,  the action~\eqref{quadphi} can be reduced to the form~\eqref{quadphiMV}.  The solution of the associated equations of motion in Fourier space is given by the sum of a vacuum homogeneous solution $u^{(0)}$  including the effect of the non-minimal kinetic coupling and a particular solution $u^{(\rm{s})}$ sourced by the axial coupling, i.e.  $u(k,\,\tau) = u^{(0)}(k,\,\tau)  + u^{(\rm{s})}(k,\,\tau)$. 

The spectrum of vacuum scalar perturbations $u^{(0)}(k)$ is computed by solving the homogeneous part of Eq.~\eqref{usourced}. To this end, we work within the approximation in which the perturbations' speed of sound $c_s$ is constant and assume a nearly de Sitter background $a \simeq -(H\, \tau (1-\epsilon_H))^{-1}$, with $\epsilon_H =-\dot{H}/H^2$. With this, we obtain
\begin{equation}
\frac{z''}{z}
\simeq  \frac{2}{\tau^2} \left[1+\frac{3}{2}\epsilon_{H} +\delta_{K} \right] +  {\cal O}(\epsilon_H^2)\,, 
\end{equation}
with $\delta_K$ defined in Eq.~\eqref{nudef}, and $z\simeq a \sqrt{K}$. Therefore, the homogeneous part of Eq.~\eqref{usourced} becomes
\begin{equation}
u^{(0)}{''}  +   \left[c_s^2\, k^2  -  \frac{1}{ \tau^2}\left(\nu^2-\frac14\right) \right] u^{(0)} =0\,, 
\end{equation}
with $\nu$ given in Eq.~\eqref{nudef}. The solution that matches the Bunch-Davies vacuum initial condition $
\lim_{\tau\rightarrow - \infty} u(k,\,\tau)  = e^{-i\, c_s\, k\, \tau}/\sqrt{2c_s\, k}$
is given by
\begin{equation}
u^{(0)} = \sqrt{\frac{\pi}{2}} \frac{e^{ i\frac{\pi}{2}(\nu + \frac12) } }{ \sqrt{2 c_s\, k  }  }\sqrt{-c_s\, k\, \tau }\, H^{(1)}_{\nu}(-c_s\, k\, \tau)\,,
\end{equation}
with $H_\nu^{(1)}$ the Hankel function of the first kind.
The super-horizon limit  of this solution 
\begin{equation}\label{u0SH}
\lim_{|c_s k\tau | \rightarrow 0}u^{(0)} (k) = \frac{2^{\nu-3/2}\,e^{i\frac{\pi}{2}(\nu - 1/2) } }{\sqrt{2 c_s\, k  }}\,\frac{\Gamma(\nu)}{ \Gamma (3/2)}\, \left(- c_s\, k\, \tau\right) ^{1/2-\nu}
\end{equation}
allows to compute the spectrum of the  vacuum primordial curvature perturbations $\zeta = -H \delta\phi/\dot{\phi}_0 = -H\, u/(z\,\dot\phi_0)$, namely
\begin{equation}
\delta(\vec{k} + \vec{k}')\,P^{(0)}_{\zeta}(k) = \frac{k^3}{2 \pi^2} \frac{H^2}{\dot{\phi}_0^2} \langle \delta \phi^{(0)}(\vec{k})\,\delta \phi^{(0)}(\vec{k}') \rangle = \frac{k^3}{2 \pi^2} \frac{H^2}{z^2\dot{\phi}_0^2}   \langle u^{(0)}(\vec{k})\,u^{(0)}(\vec{k}') \rangle\,,
\end{equation}
which, in a de Sitter background,  becomes \begin{equation}\label{vacuumPS}
P^{(0)}_{\zeta}(k) = \frac{k^3}{\pi^2} \frac{H^2}{z^2\,\dot{\phi}_0^2}  \frac{2^{2\nu-5}}{c_s\,k} \left|\frac{\Gamma(\nu)}{\Gamma(\frac32)}\right|^2 \left(-c_s\, k\, \tau\right)^{1-2\nu} \simeq  \frac{H^4}{8\pi^2\dot{\phi}_0^2} \frac{\epsilon_K}{F^2\, G\, c_s^3} \left|\frac{\Gamma(\nu)}{\Gamma(\frac32)}\right|^2 \left(\frac{-c_s\, k\, \tau}{2}\right)^{3-2\nu}.
\end{equation}

In an analogous way, we can compute the spectrum of the perturbations sourced by the axial coupling, i.e. $u^{(\rm s)}(k)$. Starting again with Eq.~\eqref{usourced} in a nearly de Sitter background, we can write
\begin{equation}\label{eq:perturbaxial}
u^{({\rm s})}{''}  +  \left[  c_s^2\, k^2 -  \frac{1}{ \tau^2}\left(\nu^2-\frac14\right)\right]   u^{(\rm{s})} = \frac{\alpha}{f}\,  \frac{a^4}{z}\, \delta [ \vec{E}_a \cdot  \vec{B}_a ]\,,
\end{equation}
with $\delta [{\vec{E}_a} \cdot  {\vec{B}_a}]\simeq \delta_{ {\vec{E}_a} \cdot  {\vec{B}_a} } + (\partial  \langle \vec{E}_a\cdot \vec{B}_a \rangle/\partial \dot{\phi} ) \delta  \dot{\phi} $. Following Ref.~\cite{Anber:2009ua},   we evaluate the variation of $\langle \vec{E}_a\cdot \vec{B}_a \rangle$ as 
\begin{equation}
 \frac{\partial  \langle \vec{E}_a\cdot \vec{B}_a \rangle }{\partial \dot{\phi} } \delta  \dot{\phi} \simeq  \frac{\partial  \langle \vec{E}_a\cdot \vec{B}_a \rangle }{\partial \xi }  \frac{\partial  \xi }{\partial \dot{\phi} }  \delta \dot{\phi} \simeq  \frac{\partial  \langle \vec{E}_a\cdot \vec{B}_a \rangle }{\partial \xi }  \frac{\alpha }{2 f H }  \delta \dot{\phi} \simeq \frac{\pi\alpha}{f} \langle \vec{E}_a\cdot \vec{B}_a \rangle \frac{\delta \phi '}{aH} \,.
\end{equation}
In the strong axial regime, we can approximate $V_{\phi} \simeq (\alpha/f)\langle \vec{E}_a\cdot \vec{B}_a \rangle $ and write the variation of the source term  as $\delta [{\vec{E}_a} \cdot  {\vec{B}_a}]\simeq \delta_{ {\vec{E}_a} \cdot  {\vec{B}_a} } + \pi V_{\phi}     \delta \phi '/(aH)$.
Taking into account that $u=z\,\delta \phi$, and neglecting the subdominant gradient term at super horizon scales, Eq.~\eqref{eq:perturbaxial} becomes
\begin{equation}\label{sourcedu}
u^{({\rm s})}{''}  +\frac{\sigma}{\tau}\, u^{({\rm s})}{'} -  \frac{1}{ \tau^2} \left(\nu^2-\frac14-\sigma   \right) u^{(\rm{s})} = \frac{\alpha}{f}\,  \frac{a^4}{z}\, \delta_{ {\vec{E}_a} \cdot  {\vec{B}_a} }\,,
\end{equation}
with $\sigma$ given in Eq.~\eqref{nuDelta}. Using now Eq.~\eqref{Apmsol} and taking into account that $u^{(\rm{s})} \simeq a\,  \sqrt{K}\, \delta \phi^{(\rm s)}$, we obtain~\cite{Anber:2009ua, Almeida:2018pir}
\begin{eqnarray}\label{psapproxphi}
\langle \delta \phi^{(\rm s)}(\vec{k})\, \delta \phi^{(\rm s)}(\vec{k}\,') \rangle 
	 & \simeq &  \frac{{\cal F} \,  H^4}{{\cal N}}   \frac{\delta(\vec{k} + \vec{k}\,')}{k^3} \left( \frac{\alpha \, {\cal N}}{\Delta \, K\,  f}\right)^2 \, \frac{e^{4\pi \xi}}{\xi^8 }  { (-2^5 \xi\, k\, \tau)^{2+2\nu_\pm}},
\end{eqnarray}
\noindent 
with ${\cal F}\simeq 2.13 \times 10^{-6}$, $\Delta$ given in Eq.~\eqref{nuDelta} and the indices $\nu_{\pm} \equiv \frac{1}{2} \left( 1- \sigma  \pm \Delta \right)$ corresponding to the growing and decaying solutions of Eq.~\eqref{sourcedu}. The $1/{\cal N}$ factor in this expression comes from assuming that the contributions of the ${\cal N}$ gauge fields  to the two-point function of $\delta_{\vec{E}_a\cdot \vec{B}_a}$ add incoherently \cite{Anber:2009ua, Anber:2012du}.
Using this result, the sourced contribution to the spectrum of  the primordial curvature perturbations $\zeta^{(\rm s)} = -H\, \delta \phi^{(\rm s)}/\dot{\phi}_0$, 
\begin{equation}
\delta(\vec{k} + \vec{k}\,')\,P^{(\rm s)}_{\zeta}(k) \equiv \frac{k^3}{2 \pi^2}\, \frac{H^2}{\dot{\phi}_0^2}\, \langle \delta \phi^{(\rm s)}(\vec{k}) \, \delta \phi^{(\rm s)}(\vec{k}\,') \rangle\,,
\end{equation}
becomes
 \begin{equation}
P^{(\rm s)}_{\zeta}(k)  \simeq  \frac{\, H^4 }{4 \pi^2 \dot{\phi}_0^2} \left[ \frac{2{\cal F}}{{\cal N}}   \left(\frac{ \alpha\, {{\cal N}} \,  H}{ \Delta \, K  \,f }\right)^2\,  \frac{e^{4\pi \xi}}{\xi^8}  { (-2^5 \xi\, k\, \tau)^{2+2\nu_\pm}}\right]\,.
\end{equation}
Combining this result  with the vacuum contribution~\eqref{vacuumPS}, we obtain the total spectrum of primordial density fluctuations in Eqs.~\eqref{eq:spesou} and \eqref{Amplitude}. These expressions are accurate up to ${\cal O}(1)$ corrections associated with the precise choice of the pivot scale $k_{*}$. 

\section{Slow-roll regime and backreaction}\label{app:slowroll}

In this Appendix we analyze the conditions allowing for an inflationary epoch  in the presence of gauge modes production and gravitationally-induced friction. To this end, we note that the first Friedmann equation~\eqref{Friedman1} can be written as the \textit{cosmic sum rule} $\Omega_\text{EM}+\Omega_{\rm K} +\Omega_{\rm V}=1$, with 
\begin{equation}\label{eq:deltaEM}
    \Omega_{\text{EM}}\equiv\frac{\langle \vec E_a^2\rangle+\langle \vec B_a^2\rangle}{6 M_P^2 H^2}\,,\hspace{15mm}     \Omega_{\rm K} \equiv \frac{\dot\phi^2}{6 M^2_P H^2}\left(1+9\frac{H^2}{M^2}\right) \,, \hspace{15mm}   \Omega_{\rm V} \equiv \frac{V(\phi)}{3M_P^2 H^2}
\end{equation}
the density parameters for the gauge fields and the inflaton kinetic and potential components. In order to have a potential-driven inflationary epoch, we need to make sure that both $\Omega_{\rm K}$ and $\Omega_\text{EM}$ are much smaller than the potential term $\Omega_{\rm V}$. More generically, the requirements $\Omega_\text{EM}\leq 1$ and $\Omega_{\rm K} \leq 1$ are consistency checks on the parameter space:
   \begin{figure}
       \centering
       \includegraphics[scale=0.6]{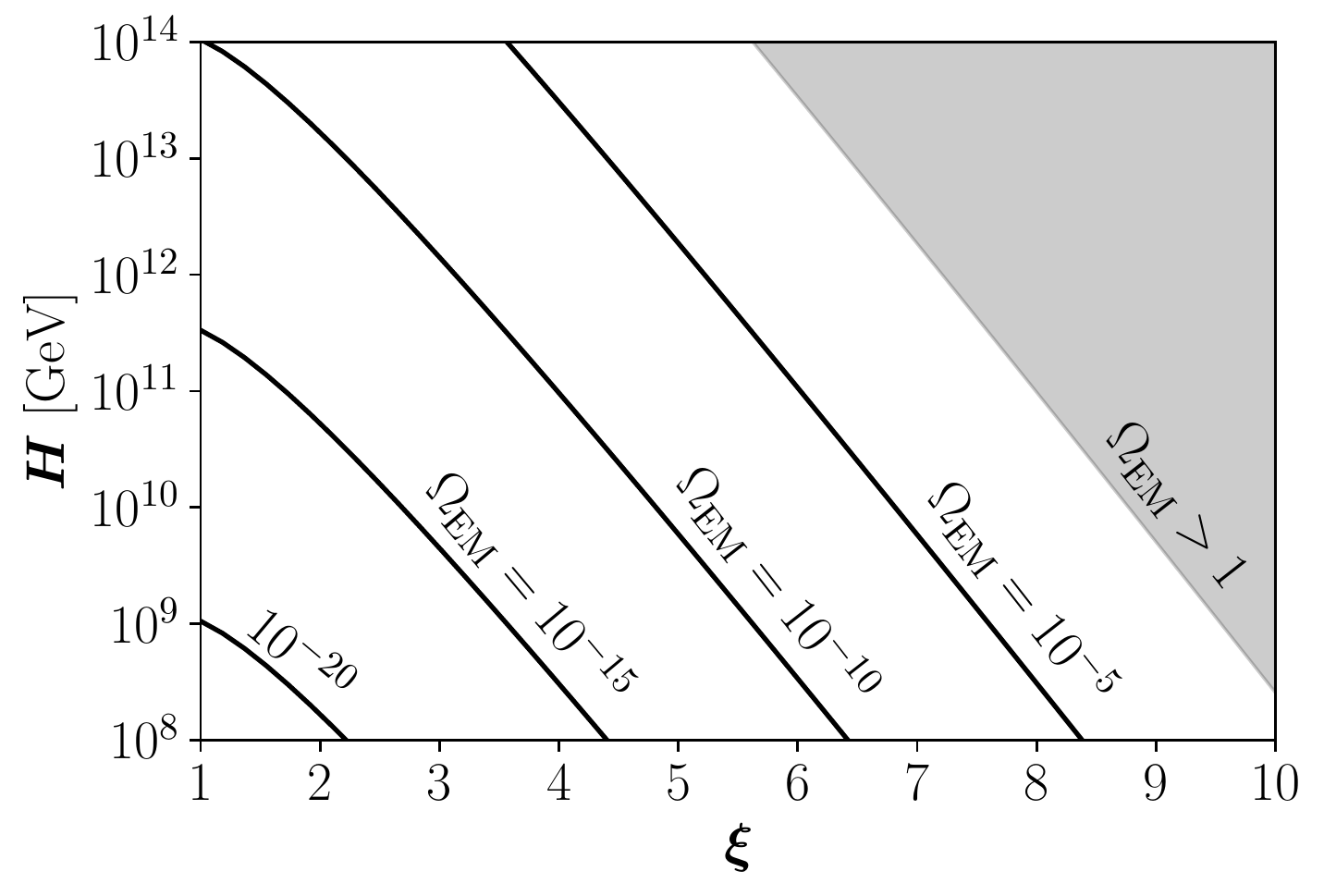}
       \caption{Level curves for the $\Omega_\text{EM}$ contribution to the energy density. Notice that the region above the curve corresponding to $\Omega_\text{EM}=1$ is forbidden as it would imply more than $100\%$ energy density stored in the gauge fields.}
       \label{fig:deltaEM}
   \end{figure} 
\begin{enumerate}
    \item Using the relations \eqref{eq:E2} and \eqref{eq:B2}, the condition $\Omega_\text{EM}\leq 1$ becomes
\begin{equation}\label{eq:epsEM}
    \Omega_{\text{EM}}\simeq3\times 10^{-16} \mathcal{N}\,e^{2\pi \xi}\left(\frac{2.6}{\xi^3}+\frac{3}{\xi^5}\right)\left(\frac{H}{10^{13}\,{\rm GeV}}\right)^2\leq 1\,,
\end{equation}
meaning that, as shown in Fig.~\ref{fig:deltaEM}, there exists a maximum value for the Hubble rate $H$ for each value of the instability parameter $\xi$. For instance, for $\xi=6$ and a single gauge field $\mathcal{N}=1$, we have $H\lesssim 10^{14}$~GeV. From a dynamical point of view, the growth of the instability parameter $\xi$ towards the end of inflation, increases the energy density of gauge fluctuations while dissipating the energy density of the inflaton condensate. This is a very efficient heating mechanism leading potentially to a very rapid thermalization for non-Abelian gauge sectors \cite{Kurkela:2011ti}.
\item Using Eq.~\eqref{eq:inst_par}, the condition $\Omega_{\rm K} \leq 1$ in the strong friction limit $H\gg M$ can be written as
\begin{equation}\label{OmegaK} 
    \Omega_{\rm K}
    \simeq6\left(\frac{\xi}{\alpha}\right)^2\left(\frac{f}{M}\right)^2 \left(\frac{H}{M_P}\right)^2\leq 1\,.
\end{equation}
Numerically, and for the range of parameters considered in this paper, this translates into an approximate relation $\alpha\gtrsim 7.5 \,\xi$.
\end{enumerate}

For the sake of completeness, we discuss also here the interplay between the gravitationally-enhanced friction generated by the non-minimal derivative coupling to gravity and the one induced by the exponential growth of gauge fluctuations. Depending on the hierarchy of scales, any of these  two independent mechanisms can a priori dominate. Here we want to determine when the contribution coming from the gauge fields becomes relevant. Assuming as usual a small acceleration in the Klein-Gordon equation \eqref{eq:phi}, the evolution of the scalar field is approximately given by $3 \,H\,K\,\dot\phi + V_{\phi} \simeq \frac{\alpha}{f}\langle \vec E_a\cdot\vec B_a\rangle$, 
where we have neglected a term proportional to $\epsilon_H \ll1$.
 Comparing the two friction terms in this equation, we get
\begin{equation}
R \equiv \left\vert \frac{\alpha\,\langle E_a\cdot B_a\rangle}{3 H K \dot\phi f}\right\vert \simeq  \frac{I_3 \,\alpha^2 \mathcal N H^2}{6 f^2 K}\frac{e^{2\pi\xi}}{\xi^5}\simeq 10^{-5}{\mathcal N}\left(\alpha \frac{M}{f}\right)^2\frac{e^{2\pi\xi}}{\xi^5}\,,
\end{equation}
where in the last equality we have assumed the high gravitational friction limit $H/M\gg1$. As long as this ratio is much smaller than one, the contribution coming from the gauge fields in the Klein-Gordon equation can be safely neglected.
For the parameter space compatible with Planck results on the amplitude and tilt of primordial density fluctuations, and assuming $N_*=60$ and ${\cal N}=20$ (left panel of Fig.~\ref{fig:numevolP}), $R$ approaches unity when the instability parameter reaches $\xi\sim 5-6$.

\bibliography{biblio} 

\end{document}